\documentclass[3p,twocolumn]{elsarticle}

\usepackage{subcaption} %Package for subfigures
\usepackage{graphicx} 
\usepackage{circuitikz} % drawings and circuits
\tikzset{>=latex} % redefine default arrow
\usepackage{amsmath} % math symbols
\usepackage{amssymb}
\usepackage{cuted}
\usepackage{changepage}
\usepackage[linesnumbered,ruled]{algorithm2e}
\usepackage{graphicx}

\newcommand{\simm}{\raise.17ex\hbox{$\scriptstyle\sim$}}
\newcommand{\bmat}{\begin{bmatrix}}
\newcommand{\emat}{\end{bmatrix}}
\newcommand{\setof}[1]{\left \{ #1 \right \}}

\newcommand {\R}{\mathbb{R}}

\begin{document}
\begin{frontmatter}

\title{Protecting residential electrical panels and service through model predictive control: A field study}

\author[purdue]{Elias N. Pergantis}
\author[purdue]{Levi D. Reyes Premer}
\author[purdue]{Alex H. Lee}
\author[purdue]{Priyadarshan}
\author[purdue]{Haotian Liu}
\author[purdue]{Eckhard A. Groll}
\author[purdue]{Davide Ziviani}
\author[purdue]{Kevin J. Kircher \corref{correspondent}}

\address[purdue]{Center for High Performance Buildings, Purdue University, 177 S Russell St, West Lafayette, IN 47907, USA}
\cortext[correspondent]{Corresponding author: \texttt{kircher@purdue.edu}}

\begin{abstract}

Residential electrification -- replacing fossil-fueled appliances and vehicles with electric machines -- can significantly reduce greenhouse gas emissions and air pollution. However, installing electric appliances or vehicle charging in a residential building can sharply increase its current draws. In older housing, high current draws can jeopardize electrical infrastructure, such as circuit breaker panels or electrical service (the wires that connect a building to the distribution grid). Upgrading electrical infrastructure often entails long delays and high costs, so poses a significant barrier to electrification. This paper develops and field-tests a control system that avoids the need for electrical upgrades by maintaining an electrified home's total current draw within the safe limits of its existing panel and service. In the proposed control architecture, a high-level controller plans device set-points over a rolling prediction horizon, while a low-level controller monitors real-time conditions and ramps down devices if necessary. The control system was tested in an occupied, fully electrified single-family house with code-minimum insulation, an air-to-air heat pump and backup resistance heat for space conditioning, a resistance water heater, and a plug-in hybrid electric vehicle with Level I (1.8 kW) charging. The field tests spanned 31 winter days with outdoor temperatures as low as -20 $^\circ$C. The control system maintained the whole-home current within the safe limits of electrical panels and service rated at 100 A, a common rating for older houses in North America, by adjusting only the temperature set-points of the heat pump and water heater. Simulations suggest that the same 100 A limit could accommodate a second electric vehicle with Level II (11.5 kW) charging. If codes permit, the proposed control system could allow older homes to safely electrify without upgrading electrical panels or service, saving a typical household on the order of \$2,000 to \$10,000. 

\end{abstract}


\end{frontmatter}

\section{Introduction}

\subsection{Electrification, infrastructure, and control}

Many people use fossil-fueled machines for driving, space heating, water heating, cooking, or drying clothes. Switching the fuel source for these activities to electricity can significantly reduce greenhouse gas emissions and air pollution in much of the world \cite{IEAbuildings, ACEEE}. However, electrifying these activities would sharply increase electricity demand, potentially straining electrical infrastructure in buildings (e.g., panels and wiring) \cite{zhou2021electrification} or power grids (e.g., transformers and power lines) \cite{sharma2021major, horowitz2019distribution}. Failure to upgrade strained electrical infrastructure can pose safety risks due to conductors overheating. On the other hand, upgrading electrical infrastructure can be slow and expensive. This paper investigates a third option: Coordinating device operation to avoid straining electrical infrastructure.

This paper focuses on circuit breaker panels and electrical service (the wires connecting a building to the distribution grid). Older buildings often have legacy panels and service sized to meet current draws from minor plug loads, but not from major appliances such as electric water heaters (WHs), heat pumps (HPs), or electric vehicle (EV) chargers \cite{nationalCode}. Electrical upgrades to accommodate major appliances require hardware and electrician labor, so can cost \$2,000 to \$10,000 for households in the United States (U.S.) \cite{DOE, FLORESETAL, WALDETAL}. 

This paper investigates whether device coordination can enable full electrification of older homes without upgrading electrical panels or service.  As proof of concept, this paper develops and field-tests a control system focused on maintaining current draws below the safe limits of the existing panel and service. The field test site is a fully electrified, occupied, detached single-family house in a cold climate. The field tests spanned 31 winter days, including a cold snap when outdoor temperatures dropped as low as -20$^\circ$C. The test house nominally requires 200 A of panel capacity under electrical codes \cite{nationalCode}, but the control system reliably maintained its current draws within the safe region of a 100 A panel, even during a cold snap when electricity demand for space heating peaked. Supporting simulations suggest that the same 100 A infrastructure could accommodate a second EV with Level II (11.5 kW) charging. The control system adjusted the temperature set-points of the HP and WH via Application Programming Interfaces that come standard with the manufacturers' communicating thermostats. The control system did not have manufacturer-level access to hardware and did not alter device-level control logic. All other appliances and Level I (1.8 kW) plug-in hybrid EV charging remained under occupant control. 

\begin{figure}
\centering
\includegraphics[width=0.475\textwidth]{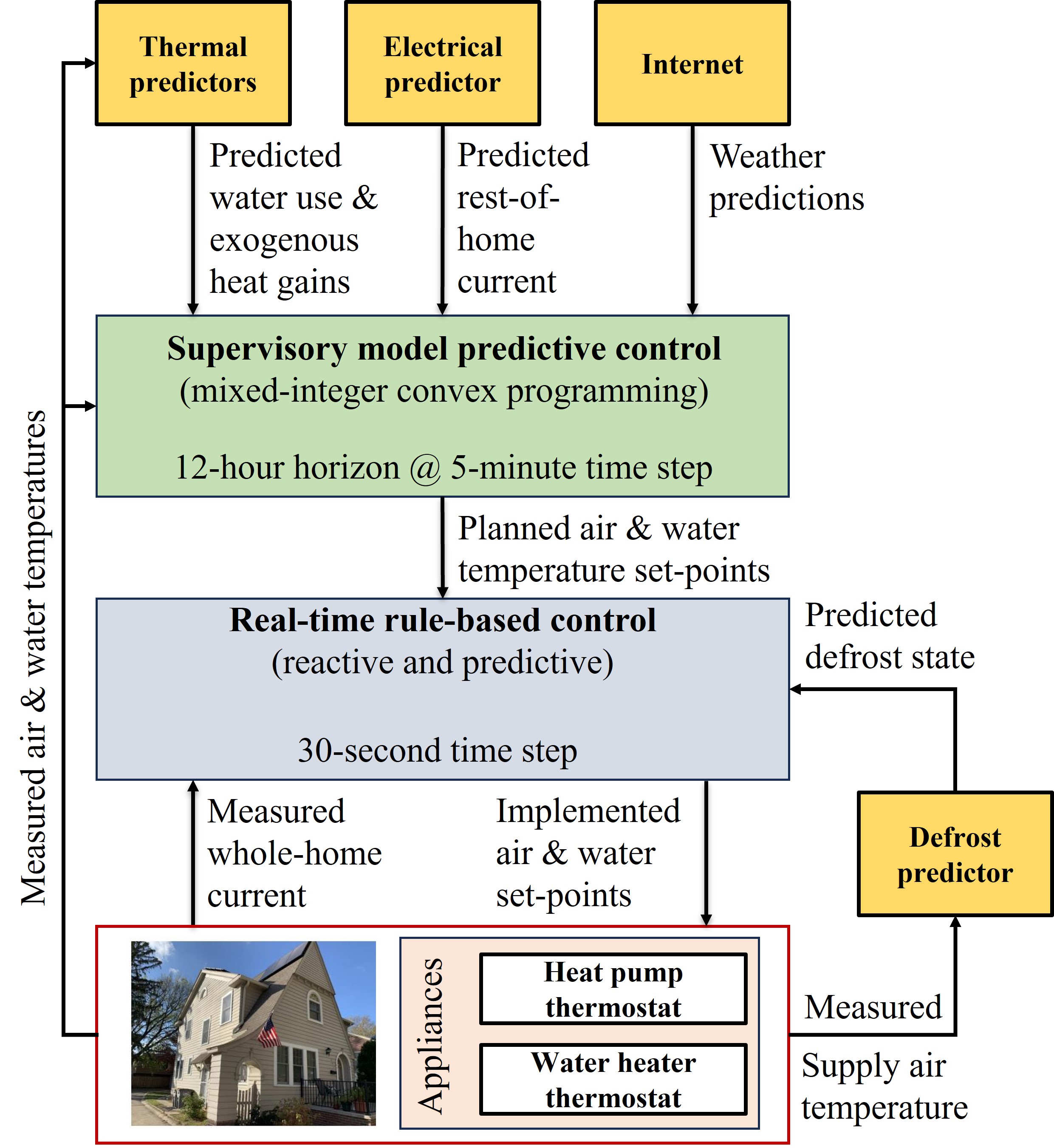}
\caption{The control system uses real-time measurements from the house to decide HP and WH set-points that maintain comfort and ensure safe whole-home currents.}
    \label{flowchart}
\end{figure}

\begin{figure*}
\centering
\begin{subfigure}[t]{0.45\textwidth}
    \includegraphics[height=5cm]{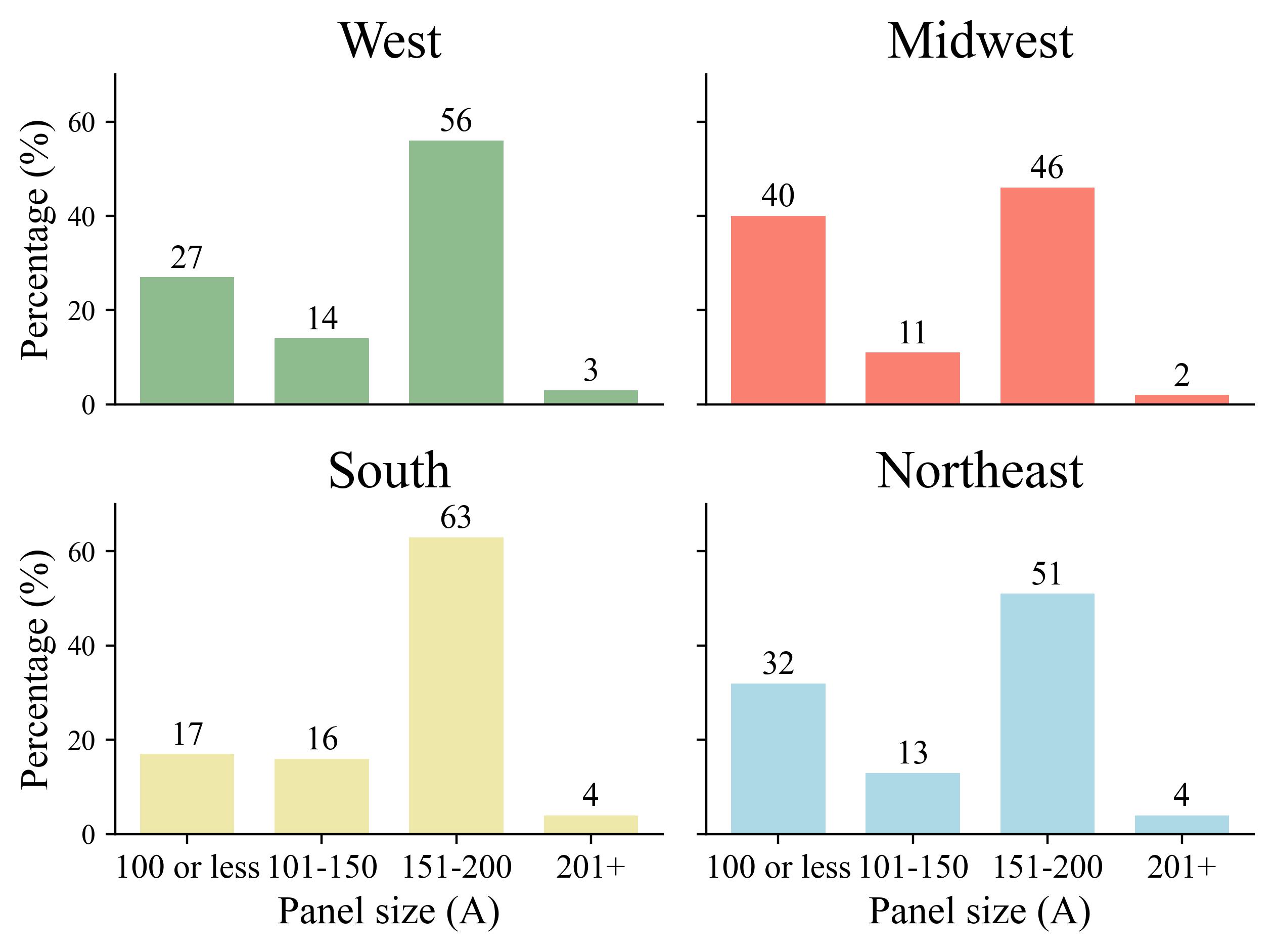}
    \end{subfigure}
\begin{subfigure}[t]{0.5\textwidth}
    \includegraphics[height=4.7cm]{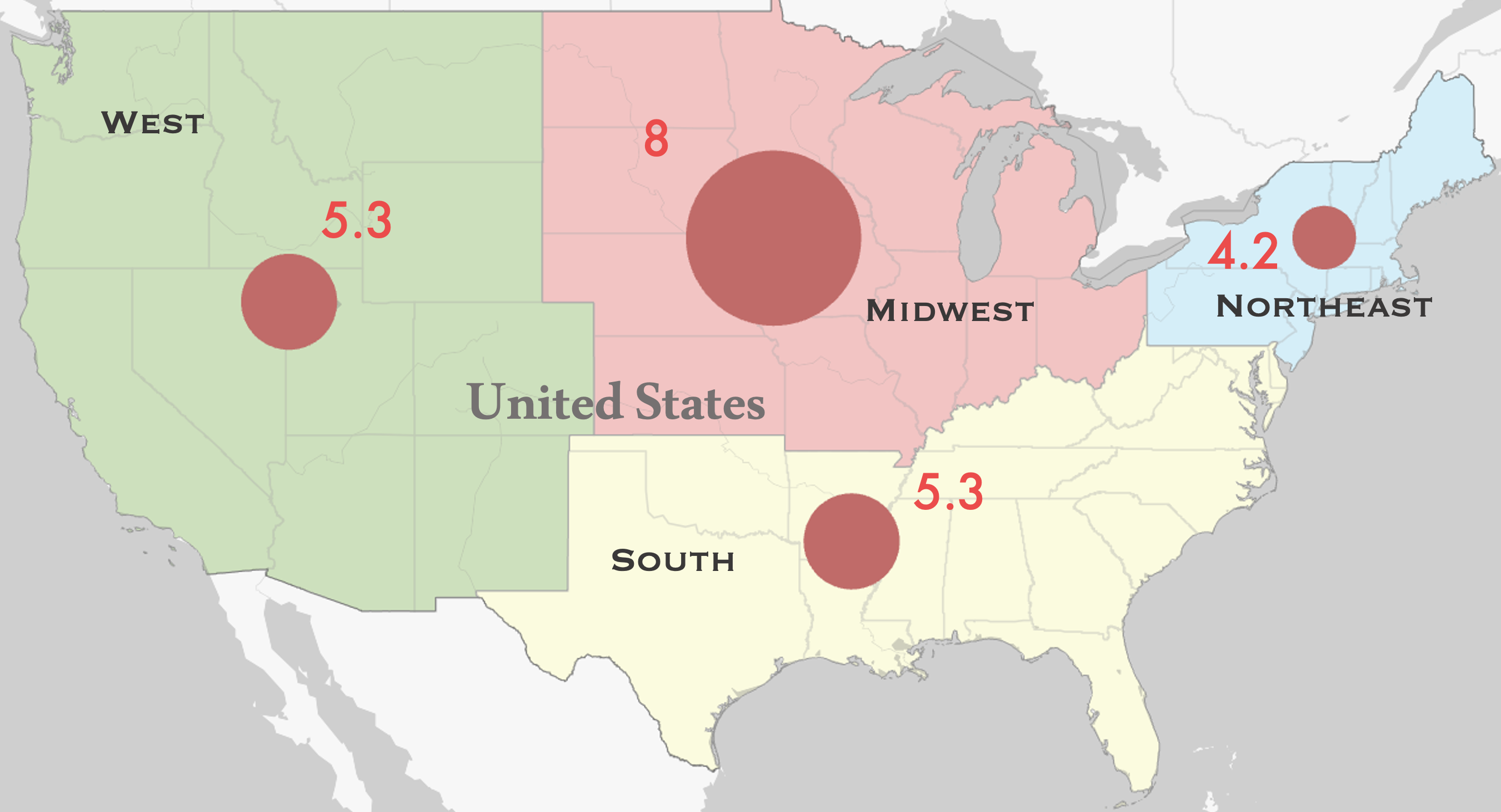}
    \end{subfigure}
\caption{(Left) Percentage breakdown of panel sizes in four census regions \cite{EPRI}. (Right) Number of panels (in millions) that require an upgrade in the case of full electrification with backup heat and electrical vehicle adoption.}
\label{panel_by_region}
\end{figure*}

\subsection{Contributions of this paper}
\label{contributionsSection}

This paper makes four main contributions.
\begin{enumerate}
    \item It investigates the under-studied problem of protecting or upgrading electrical panels and service in the context of residential electrification retrofits \cite{DOE,LA100, FLORESETAL, LESSETAL}. To frame the problem, this paper reviews electrical panel sizes in the U.S. and upgrade needs under deep electrification scenarios. This paper discusses safe operating regions for electrical panels, analyzing the test house as a case study, then reviews existing literature on protecting electrical infrastructure through device coordination. 
    
    \item This paper develops a novel two-level control architecture to address the identified problem. In this architecture, shown in Fig. \ref{flowchart}, the high-level controller plans device set-points over a receding forecast horizon using scenario-based model predictive control (MPC) and mixed-integer convex programming. The low-level controller uses rule-based logic, turning devices off in stages if real-time measurements of the whole-home current approach unsafe levels. The low-level controller also predicts the use of resistance backup heat during HP defrost cycles, a dominant driver of electricity demand peaks, by monitoring supply air temperatures. 
    
    \item As far as the authors are aware, this paper reports the first field demonstration of whole-home current-limiting control in the research literature \cite{Arash_review}. The experiments demonstrate the potential of advanced control, using smart meters and thermostats, to reliably maintain current draws in fully electrified homes within the safe limits of legacy electrical panels and service, even in cold climates.
    
    \item This paper develops a unified approach to learning data-driven, control-oriented models that capture the thermal and electrical behavior of both WHs and HPs. Further, this paper presents the first field demonstration of a predictive control system that coordinates multiple devices within in a residential building to achieve a common objective. The potential of this has been hypothesized in simulation \cite{7114268, GASSER2021116653,SALPAKARI2016425} but, to the best of the authors' knowledge, has not been tested in a real, occupied residence. 

\end{enumerate}

This paper expands upon the first field results presented in \cite{pergantis2024herrick}, adding deeper discussion of related literature, further details on the controller design methodology, new experiment results, and new simulation results that include a second EV.

Section \ref{background} of this paper discusses the problem, with the test house as a case study. Section \ref{high_level_formulation} presents the high-level controller, including models of the building, HP, WH, and the rest-of-home current. Section \ref{low_level_controller} presents the low-level controller and defrost prediction methodology. Section \ref{testing_results} presents field experiment results, as well as simulations that extend the control system to incorporate a second EV with Level II charging. Section \ref{future_work} discusses practical considerations and future work. The appendix defines the acronyms and mathematical notation used in this paper.

\section{Understanding the problem}
\label{background}

This section introduces the problems that can arise with electrical panels and service when electrifying residential buildings. It begins with an overview of panel sizes in U.S. housing, then discusses panel failure mechanisms, drawing on historical data from the test house as a case study. This section then formalizes the current-limiting control problem and contextualizes this paper's contributions relative to existing literature.

\subsection{Circuit breaker panel sizes in the U.S.}

One strategy to meet greenhouse gas emission goals is to electrify residential buildings. To this end, many U.S. states have policy objectives for residential electrification. For example, both the California Senate Bill 1477 and the New York State Clean Heat program strongly support the adoption of HPs \cite{ACEEE}. However, a significant barrier to residential electrification is the current state of electrical infrastructure in the U.S. Major appliances may require electrical panel or service upgrades, which may exceed the cost of the appliances themselves. These upgrades may not always be financially viable, especially for low-income households. This section focuses specifically on circuit breaker panels, with the understanding that electrical service capacities match panel capacities reasonably closely in most homes.

\begin{figure*}
\centering
\begin{subfigure}{0.32\textwidth}
    \includegraphics[width=\textwidth]{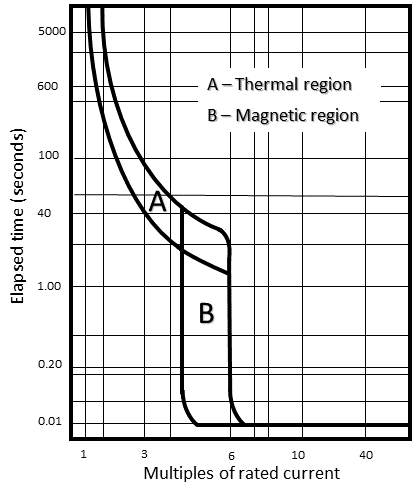}
    \end{subfigure}
\begin{subfigure}{0.45\textwidth}
    \includegraphics[width=\textwidth]{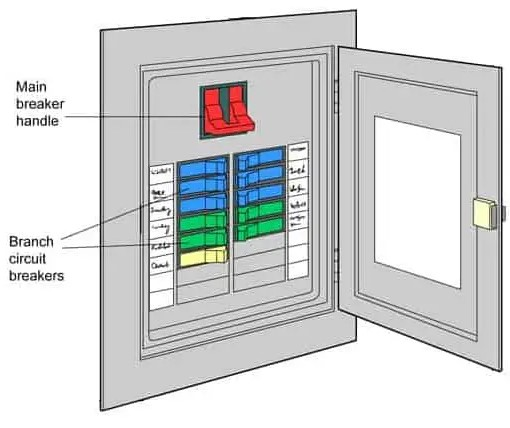}
    \end{subfigure}
\caption{(Left) Different magnitudes and durations of current peaks trigger thermal vs. magnetic trips in circuit breakers. (Right) A typical panel has a main breaker and several branch circuit breakers \cite{breakerFig}. This paper focuses on the main breaker.}
\label{Breaker&TripCurve}
\end{figure*}

Few studies exist that characterize the challenge of upgrading residential electrical infrastructure in the context of electrification retrofits \cite{ACEEE, DOE, LA100, FLORESETAL, LESSETAL}. To the best of the authors' knowledge, there are limited nation-level data on panel sizes. A 2023 report from the U.S. Department of Energy's Office of Energy Efficiency \& Renewable Energy \cite{DOE} set out to create publicly available resources to provide electrification solutions under electrical panel and service constraints. They estimated that about 21\% of U.S. homes have 100 A or less panel capacity and about 44\% of homes have two or fewer slots available in their panels. One report from the Los Angeles Department of Water \& Power estimates that about 46\% of single-family homes in disadvantaged communities in Los Angeles have panels rated at 100 A or less and about 31\% between 100 A and 200 A. In total, they estimate that it will require between \$780 million and \$1.8 billion to upgrade all deficient single-family residential service panels for full electrification, just in Los Angeles \cite{LA100, FLORESETAL}. Pecan Street \cite{Pecanstreet} surveyed 263 residences in Austin, Texas. They noted that the floor area of the house, the year the house was built or last renovated, and the fuel source for home appliances have the most impact on the rated size of breaker panels. However, these factors only serve as an approximation for the actual panel size and much of this data is not publicly available (such as renovation dates). 

The same U.S. Department of Energy report also highlighted that gas-heated homes generally have panels with a smaller capacity than electrically heated homes \cite{DOE}. Coupled with the fact that colder climates require larger space heating equipment and often high-power backup resistance heating elements, the challenge of deep electrification is clear. Pecan Street \cite{Pecanstreet} also highlighted that less than 20\% of the homes in the West, Midwest, and Northeast are all-electric, compared to 44\% in the South. This finding echoes the Electric Power Research Institute's (EPRI's) recent report \cite{EPRI}, which found that of homes in the U.S. with 100 A or less, 80\% have three or fewer large electrical appliances. Fig. \ref{panel_by_region} summarizes the distribution of panel sizes in the four U.S. regions.

As noted by both Pecan Street and EPRI, the full electrification of homes will most likely require a panel upgrade of at least 200 A under today's National Electrical Code guidelines \cite{nationalCode}. EPRI provides a detailed analysis of 10 electrification scenarios and the percentage of households requiring panel upgrades \cite {EPRI}. Of interest are two scenarios: full electrification of home appliances with backup heat (as is the case in the test house in this paper), and full electrification of home appliances with backup heat and electric vehicle adoption. In the former scenario, EPRI estimates that 11.5\% of single-family homes in the U.S. will require panel upgrades. In the latter scenario, EPRI estimates that 24\% of all single-family homes will require a panel upgrade (35.8\% in the Midwest). Extrapolation of EPRI's analysis should be done with care due to the relatively small sample size. However, applying these results to the number of single-family units in each region, the number of single-family homes that could require panel upgrades in the U.S. is about 23 million. Fig. \ref{panel_by_region} shows the geographical breakdown of these estimates. With the average cost of upgrading each panel around \$2,000 to \$10,000, the total cost to upgrade electrical panels in single-family homes could well be in the tens or hundreds of billions. These estimates exclude multi-family homes, which generally have lower panel capacities per unit (50 - 70 A \cite{LESSETAL}). These estimates also exclude any further upgrades necessary to electricity generation, transmission, or distribution infrastructure.

\subsection{Safety limits and test house}
\label{problem_statement}

\subsubsection{Panel failure mechanisms}

Most panel upgrades involve installing a new panel with a higher rated current capacity. Such upgrades allow for safe operation and prevent the main circuit breaker from repeatedly tripping, which can inconvenience users and shorten the panel's lifespan \cite{9087009}. Circuit breakers prevent fires by opening circuits when currents approach levels that could overheat conductors. When a breaker trips, it prevents current flow within the circuit it governs. 

Typical breakers have two tripping mechanisms, as shown at left in Fig. \ref{Breaker&TripCurve}. The first mechanism, thermal tripping, is caused by heat generation within the breaker. As the current increases, Joule heating within conductors causes the temperature of the breaker to rise; the breaker trips when its temperature gets too high. Thermal tripping can happen after high current draws are sustained for seconds to minutes, with the timing determined by the magnitude of the current. The greater the current magnitude, the sooner the breaker will trip. The second mechanism, electromagnetic tripping, is caused by brief but very high current peaks, typically five or more times higher than the rated current. The breaker will trip within milliseconds to seconds of a severe current peak \cite{Larsen2008}. Both thermal and electromagnetic failure mechanisms can be identified in the trip curve developed by the manufacturer, which provides the range of times the breaker will trip at a certain multiplicity of the rated current. A ``safe zone'' and an ``unsafe zone''  are separated by the breaker's tripping region seen in the time vs. current curve. 

As shown at right in Fig. \ref{Breaker&TripCurve}, a typical home electrical panel contains a main circuit breaker and a number of branch circuit breakers that monitor either an individual large appliance (such as a dryer or electric heater) or circuits with multiple smaller loads (such as a bedroom or kitchen). For the purposes of this paper, the branch circuit breakers are assumed to be correctly sized and sufficient in number to accommodate the new devices in an electrification retrofit. Under this assumption, all relevant failures are within the main circuit breaker. For this reason, the rest of this document refers only to the main circuit breaker. However, the methods in this paper could be straightforwardly adapted to protect one or more branch circuit breakers instead of or in addition to the main circuit breaker.

\begin{figure}
\centering
\includegraphics[height=0.35\textwidth]{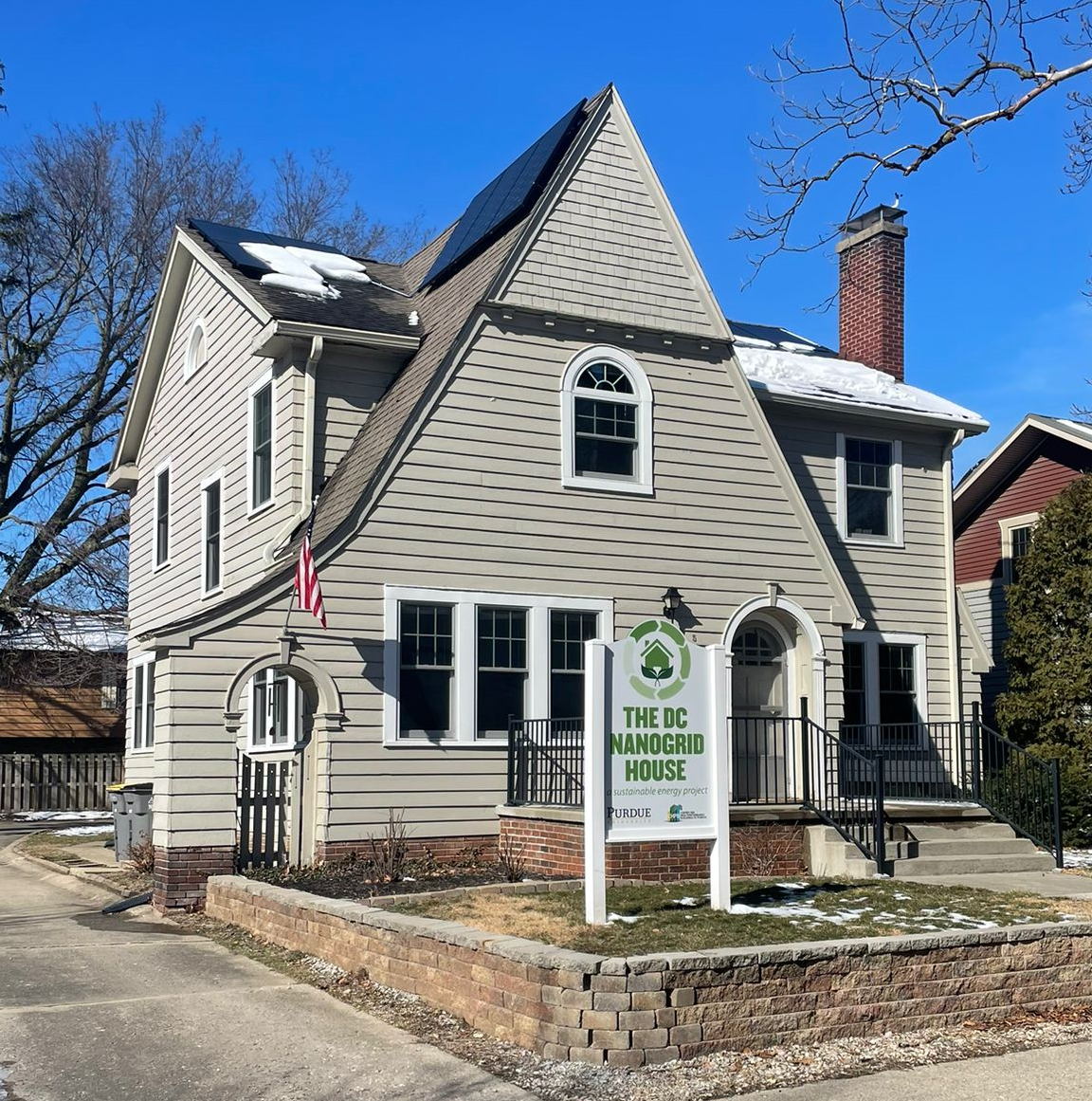}
\caption{The test house is a 208 m$^2$, 1920s-era house with all-electric appliances in West Lafayette, U.S.}
\label{DCHouseFig}
\end{figure}

\subsubsection{Test house}
\label{testHouse}

The field demonstrations in this paper took place in the 1920s-era, 208 m$^2$ detached single-family house shown in Fig. \ref{DCHouseFig}. The house is in West Lafayette, Indiana, a U.S. location that falls under the International Energy Conservation Climate Code Zone 5A \cite{IECC2021}. This climate zone sees winter temperatures as low as -20 $^\circ$C. In recent years, the house has been renovated with code-minimum insulation (R-value of 3.5 $^\circ$C m$^2$/W) and windows (U-value of 8 W/m$^2$/$^\circ$C), as well as all-electric appliances for space conditioning, water heating, cooking, laundry, etc. \cite{pergantis2024field}. Throughout the testing period, the occupants periodically charged one plug-in hybrid EV with 8.8 kWh of battery capacity and Level I (1.8 kW) charging. Simulations in Section \ref{ev_sim_section} incorporate an additional EV with a 70 kWh battery and Level II (11.5 kW) charging.

The test house is conditioned by a central ducted air-to-air HP with backup resistance heating elements. The HP has 14 kW of rated cooling capacity, a cooling seasonal coefficient of performance (COP) of 5.3, and a heating seasonal COP of 2.5. Up to 19.2 kW of backup resistance heat turns on in stages (9.6, 14.4, and 19.2 kW) when the HP's vapor-compression cycle cannot keep up with heating demand in very cold weather. The heating elements also run during defrost operation, when the HP reverses the vapor-compression cycle to melt frost build-up on the outdoor heat exchanger. 

Hot water in the test house is provided by a 50-gallon (189 L) hybrid electric WH equipped with a HP and two electric resistance heating elements. The two heating elements are at the bottom and top of the tank with electrical capacities of 4.5 kW and 2.25 kW, respectively. This enables the WH to also run in resistance-only mode, where the HP is disabled and only the resistance heating elements are used. Because resistance WHs are far more common in the U.S. than heat-pump WHs \cite{RECS2020WH}, the experiments in this paper were run with the WH in resistance-only mode. This choice also made current-limiting control problem more challenging, as resistance-only WHs draw significantly more current than heat-pump WHs.

Under the National Electric Code article 220 \cite{nationalCode}, the test house requires 200 A of rated panel capacity. Because it was renovated for research purposes, the house now has two 200 A panels, which enable connection of a large photovoltaic array, a stationary battery, a Level II electric vehicle charging station, etc. These additional features were all disabled during the current-limiting control demonstrations in this paper. A home that has not fully electrified would typically have a panel and service rated at 80 to 100 A, a common range for older U.S. homes. Therefore, the field demonstrations in this paper sought to maintain whole-home current within the safe limits of a 100 A panel and service.

\subsubsection{Baseline operation}
\label{baselineOperation}

\begin{figure}
\centering
\includegraphics[width=0.4\textwidth]{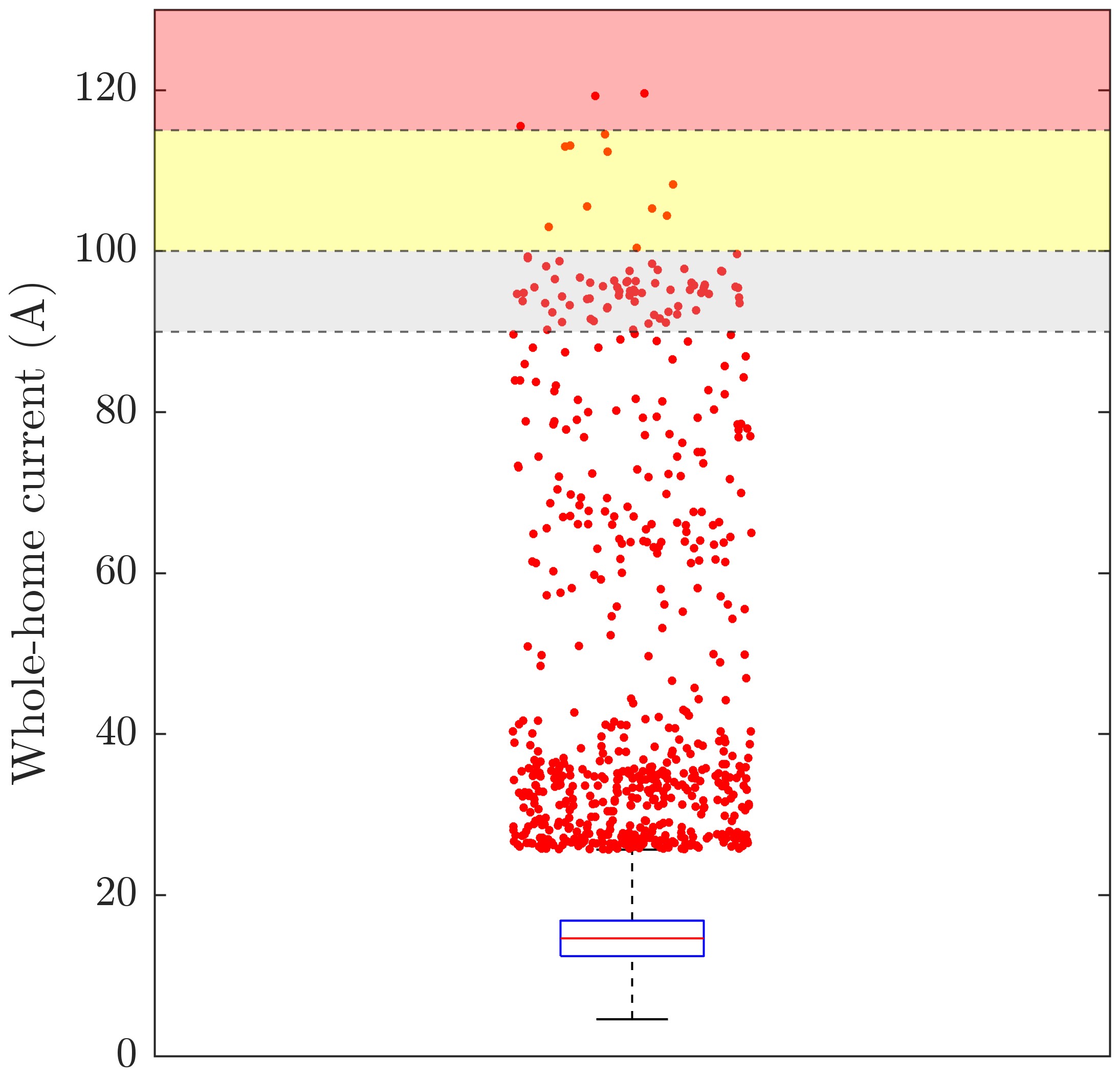}
\caption{Five-minute whole-house current over 43 baseline days. Red line: Median. Blue box: 25th to 75th percentile range. Red dots: Outliers. Without device coordination, currents were often undesirable (gray region), occasionally unsafe for older 100 A breakers (yellow), and three times would likely have tripped even a new 100 A breaker (red).}
\label{LY_Current_5min}
\end{figure}

To characterize baseline operation, the total current draw in the house during the winter of 2022-23 was analyzed. Summer operation is not considered since the test house's current peaks are driven by backup resistance heating during the coldest weather. In 2022-23, the WH ran in HP mode. To enable fair comparison between the 2022-23 baseline and the 2023-24 experiments, the WH's current draws during the baseline period were increased to emulate a resistance WH. This was achieved by propagating the 2022-23 water draws through a calibrated model of a resistance-only WH under thermostatic control. Fig. \ref{LY_Current_5min} illustrates the resulting current draw for the house with the WH adjustments. Although code requires a 200 A panel, the current surpassed 100 A only a few times. However, these few times could have tripped a 100 A breaker, shutting down power to the house. 

The shaded regions in Fig. \ref{LY_Current_5min} represent different levels of safety, as obtained from the trip regions in Fig. \ref{Breaker&TripCurve} for a typical 100 A breaker and a five-minute time step. The five-minute time step is consistent with the time scales of relevant thermal tripping mechanisms (region `A' in Fig. \ref{Breaker&TripCurve}) due to current draws of less than three multiples of the rated current, which span about one minute to one hour. The gray zone in Fig. \ref{LY_Current_5min} represents the undesirable region where the home's current draw is between 90-100 A, as this range is prone to tripping if additional large equipment switches on. The yellow region, between 100-115 A, marks the unsafe zone for the breaker over its 100 A rated current. This zone is yellow because prolonged exposure can lead to thermal tripping due to increased breaker temperatures, especially in older panels. The red region indicates where thermal tripping of the breaker is possible even in a new breaker. During the 32-day baseline period, the house's original operation had three likely failures (red region) and ten instances in the potentially unsafe (yellow) region. Table \ref{tab:plim} summarizes the numbers and durations of current limit violations in the 2022-23 baseline period.

\begin{figure}
\centering
\includegraphics[width=0.45\textwidth]{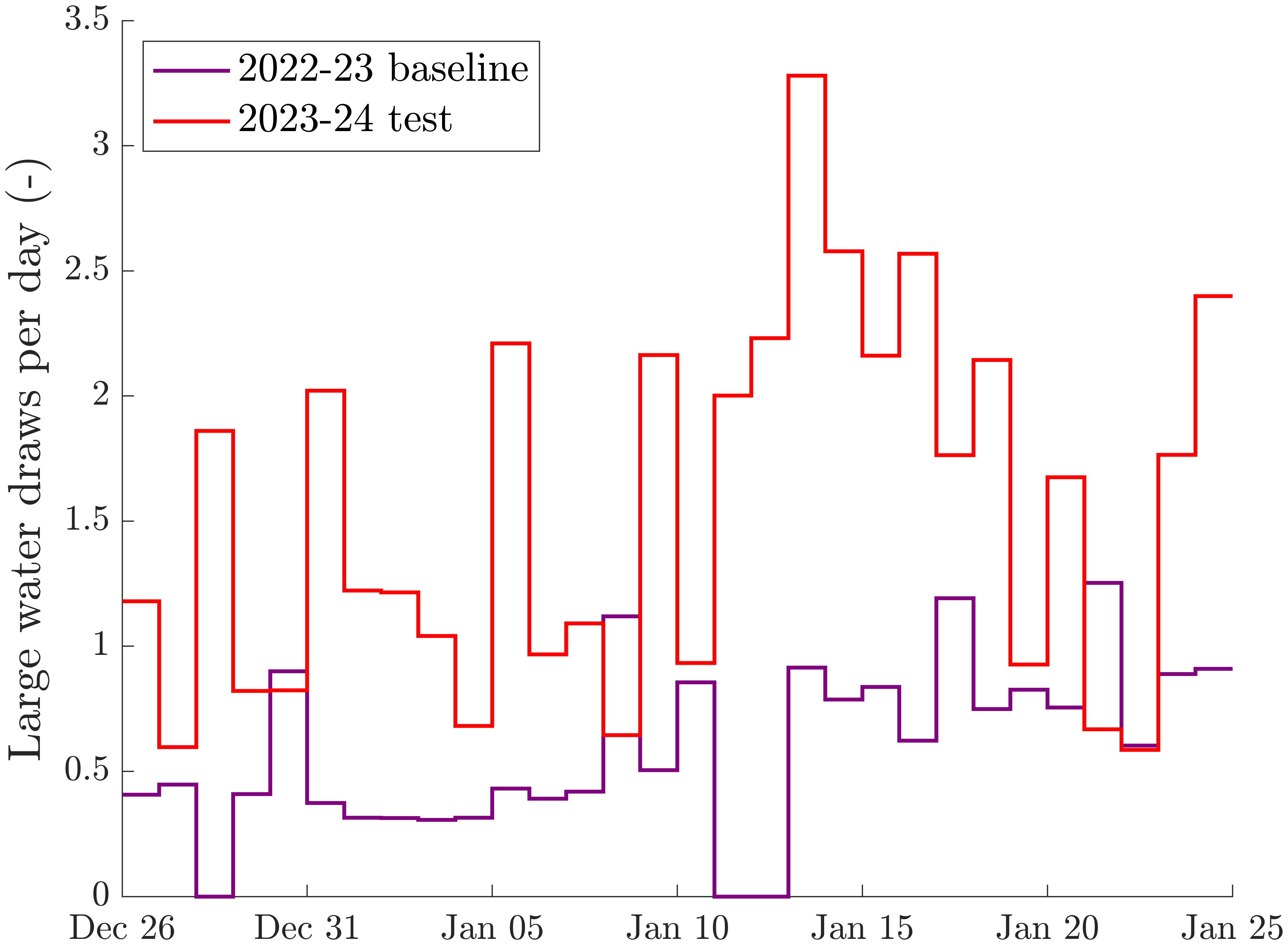}
\caption{In the 2022-23 baseline period (purple), the test house's two occupants used significantly less hot water than the three occupants used in the 2023-24 test period (red).}
\label{Hot_Water_Usage}
\end{figure}

Occupant behavior is a key driver of device synchronization events that cause current peaks. The more occupants, the more likely it is that multiple devices (such as the WH and HP) will run simultaneously. In the winter of 2022-23 (the baseline period), the test house had only two occupants. By contrast, the test house had three occupants during the winter of 2023-24 (the advanced control testing period). Fig. \ref{Hot_Water_Usage} shows how this increase in occupancy increased hot water usage in the winter of 2023-24 relative to 2022-23. In Fig. \ref{Hot_Water_Usage}, the hot water usage was normalized over the house's average shower duration of 10-12 minutes and hot water volume of $\simm$12 gallons ($\simm$45 L). The water usage in 2023-24 (red curve) was in some cases three times higher than the previous winter's usage (purple curve). This increased the likelihood of synchronization of the WH with other equipment and made the 2023-24 control problem more difficult.  

\begin{table*}[ht]
\centering
\caption {Current limit violations in the 2022-23 baseline period} \label{tab:plim} 
\resizebox{0.8\textwidth}{!}{%
\begin{tabular}{ccccc}

  \begin{tabular}[c]{@{}c@{}}Current\\ Limit (A) \end{tabular} &
  \begin{tabular}[c]{@{}c@{}}Total minutes\\ of violation\end{tabular} &
  \begin{tabular}[c]{@{}c@{}}\# violations longer\\ than 10 minutes\end{tabular} &
  \begin{tabular}[c]{@{}c@{}}Average minutes\\ of violation\end{tabular} &
  \begin{tabular}[c]{@{}c@{}}\# violations \\ per day \end{tabular} \\ \hline
80  & 730 & 54 & 8   & 3.4                  \\
90  & 480 & 19 & 6.2 & 2.2                  \\
% 95  & 255 & 6  & 5.7 & 1.2                  \\
100 & 70  & 3  & 6.4 & 0.3 (2.1 per week) \\
110 & 40  & 0  & 5   & 0.19 (1.3 per week)   \\ 
\end{tabular}%
}
\end{table*}

\subsection{Current-limiting problem structure}
\label{problemStructure}

This section describes one approach to the problem of running an all-electric house within the safe limits of electrical panels and service rated at 100 A. Trip curves for 100 A panels show that for the current profile observed in the test house (Table \ref{tab:plim}), there is negligible risk of magnetic trips due to currents exceeding 500 A. Therefore, this section focuses on the problem of avoiding thermal trips.

To clarify the problem of avoiding thermal trips, the authors analyzed trip curves for several 100 A breakers from major manufacturers. While the current draw should ideally be maintained below 100 A \cite{nationalCode}, trip curves show that 100 A breakers can handle up to 115 A for up to five minutes without tripping. Trip curves also show that current draws of 80 A sustained for three hours or more can cause thermal trips. These considerations give rise to three goals for avoiding thermal trips: Maintain the whole-home current i) below 115 A at all times; ii) below 100 A whenever possible; and iii) below 80 A under normal conditions.

In the test house, peak current draws happen during the heating season, driven primarily by the HP and WH. The HP's compressor draws up to 22 A during normal heating operation and up to 7 A during defrost cycles. The HP has three stages of backup resistance heat. Stages I, II and III draw 40, 60, and 80 A, respectively. The WH draws 18.7 A. In combination, all other devices in the house rarely draw more than 20 A and never more than 40 A (see the righthand plot in Fig. \ref{binned_current_draw}). 

Pragmatically, the current-limiting problem largely boils down to avoiding synchronization of backup heat with other major loads. By itself, an 80 A draw from the HP's highest stage of backup heat can place a 100 A panel at risk. Simultaneous use of any stage of backup heat with the compressor and WH is also risky. As the HP uses backup heat during every defrost cycle, it turns out that accurately predicting defrost cycles (see Section \ref{defrost_section}) is crucial for current-limiting control.

\subsection{Overview of related research}

Advanced control of distributed energy resources (such as batteries, EVs, rooftop photovoltaics, WHs, HPs, and other heating, ventilation and air conditioning equipment in buildings) is an extensive field of research, as reviewed in \cite{drgovna2020all, en11030631}. As far as the authors are aware, no studies have specifically demonstrated current-limiting control in residential or commercial buildings to protect electrical panels or service. However, current-limiting control is closely related to the problems of reducing and constraining peak power demand. Therefore, this section focuses on papers that reduce peak demand charges \cite{kircher_peak, peak_disaggregation, peak_disaggregation2}, provide demand response \cite{GASSER2021116653,  KIM201849}, or satisfy a hard power constraint \cite{7114268}. This section discusses past work on a) hardware-based methods, b) real-time software-based methods, and c) supervisory software-based methods. Finally, this section highlights the novel contributions of this paper in the context of past work.

Hardware-based methods for power regulation have been widely investigated \cite{micro_grids_control}. This field includes design of photovoltaic inverters \cite{currentinverter, 10362266} and other power electronics \cite{Goldin2024}. Particularly relevant are power-splitting devices, which force a circuit to power only one load at a time. A popular use of power-splitting devices is to interleave the charging of two EVs plugged into the same outlet. There have also been extensions of this work to subcircuit breakers that power multiple loads \cite{8336989}.

In the area of real-time software-based methods for reducing or constraining demand, multiple research threads exist. Some consider thermostatically controlled loads \cite{HAM2023113351} or other devices, such as EV charging \cite{7114268}. Studies use centralized \cite{peak_disaggregation} or distributed \cite{kircher_peak} control architectures. Some methods use numerical optimization \cite{9084100}, while others use rule-based heuristics \cite{6531077, barker_smartcap_2012}. Methods can also be classified as model-based \cite{7114268} or model-free \cite{kircher_peak, peak_disaggregation2}. 

In the area of supervisory software-based methods, many studies have investigated device coordination in simulation to reduce peak demand \cite{GASSER2021116653,SALPAKARI2016425,7114268, barker_smartcap_2012}. In \cite{GASSER2021116653}, coordination of an EV and HP (for space heating and domestic hot water), was simulated for demand response. Barker et al. \cite{barker_smartcap_2012} developed a least-slack-first heuristic to shift home loads, reducing peak power by 20\% in simulations with a real home's energy usage data. Papers have also proposed meta-heuristics for whole-home power limiting \cite{della_croce_heuristic_2017, awais_towards_2018}. These algorithms typically do not guarantee power constraint satisfaction, making them less than ideal for current-limiting control. Most relevant to this paper is \cite{7114268}, which incorporated detailed models for various distributed energy resources to achieve hard and soft power-limiting control in simulation. The controller in \cite{7114268} used mixed-integer optimization with a five-minute time step and 12-hour prediction horizon, similar to this paper. However, the simulations in \cite{7114268} did not include the unforeseen disturbances that can arise in real buildings and increase the risks of violating constraints.

A small number of experimental studies have demonstrated supervisory software-based methods for reducing or constraining peak demand in real buildings. Kim et al. \cite{KIM2015279, KIM201849} demonstrated the potential of coordinating the set-points of multiple air conditioners in commercial buildings to reduce peak demand. In \cite{pergantis2024humidity}, Pergantis et al. demonstrated power-limiting control for a residential air conditioner and investigated the impact of model accuracy on constraint satisfaction. 

In addition to the contributions enumerated in Section \ref{contributionsSection}, this paper makes two main methodological innovations. First, the methods developed here are designed to be implementable with little hardware cost. The control system interacts with the HP and WH over Wi-Fi via Application Programming Interfaces that manufacturers provide by default with communicating thermostats. These interfaces enable temperature sensing and set-point actuation. The control system pulls current measurements from a power meter with circuit-level monitoring; these devices cost about \$100 retail. 

Second, the methods developed here fuse fast device-level control, similar to \cite{kircher_peak}, with supervisory planning, similar to \cite{7114268}. The supervisory control methodology draws inspiration from robust \cite{robust_mixedintegerfoundations} and scenario-based \cite{1632303} methods for mixed-integer convex programming. Similar approaches have previously been investigated for utility operation and process control \cite{QIU2023121693,SHEN2020114199}, but not for residential current limiting. The supervisory control methodology also involves physics-based and machine-learning approaches to modeling building thermal dynamics, domestic hot water draws, and combined current draws from all non-thermal loads. The overall control system design works well for residential current limiting, and could find use in other applications where constraint satisfaction is a major focus.

\section{High-level controller}
\label{high_level_formulation}

This section discusses the predictive models used by the supervisory controller, including the thermal dynamics, HP, WH, and miscellaneous electrical load. This section also formulates the optimization problem that MPC solves at each time step. 

\begin{figure}[ht]
\centering
\begin{circuitikz}[scale=1.1, american currents] 
\ctikzset{bipoles/length=1.05cm} % rescale resistors and capacitors
% variables
\pgfmathsetmacro{\w}{2};
\pgfmathsetmacro{\h}{1};

% ambient > ground
\node[above] at (2*\w,2*\h) {$T_m$};
\draw (2*\w,2*\h) to[battery1,*-] (2*\w,0) -- (0,0);

% air > ambient
\node[above left] at (\w,2*\h) {$T$};
\draw (0,2*\h) to (\w,2*\h) to[R,R=$R_m$,*-] (2*\w,2*\h);

% thermal power source
\draw (0,0) to[I,n=Qa] (0, 2*\h);
\node[right] at (Qa.s) {$\dot Q + w$};

% ground > air
\draw (\w,0) to[C,n=Ca] (\w,2*\h);
\node[right] at (Ca.s) {$C$};
\draw (\w,0) node[ground] {} to (\w,0);

% air > adjacent
\draw (\w,2*\h) -- (\w,3*\h) to[R,R=$R_\text{out}$,-*] (3*\w,3*\h);% -- (3*\w,2*\h);
\draw(3*\w,3*\h) to[battery1] (3*\w,0) -- (2*\w,0);
\node[above] at (3*\w,3*\h) {$T_\text{out}$};

\end{circuitikz}
\caption{A thermal circuit with two resistors and one capacitor models the building temperature dynamics.}
\label{2R1CFig}
\end{figure}
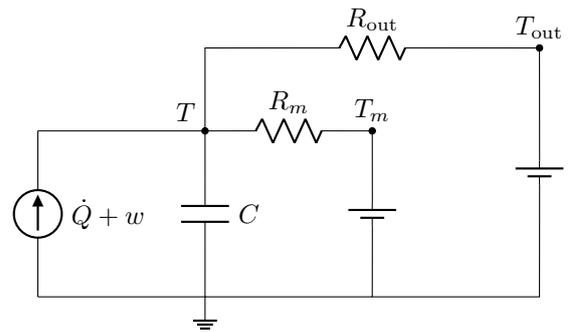

\subsection{Building envelope thermal modeling}
\label{rc_model}
\subsubsection{Thermal circuit model}

Fig. \ref{2R1CFig} shows a thermal circuit model of the test house. This thermal circuit can be understood by analogy to an electrical circuit, with thermal power playing the role of current and temperature playing the role of voltage. The state is the indoor air temperature $T$ ($^\circ$C). The parameters are the resistance $R_\text{out}$ ($^\circ$C/kW) between the indoor and outdoor air, the resistance $R_m$ ($^\circ$C/kW) between the indoor air and the interior thermal mass, the thermal mass temperature $T_m$ ($^\circ$C), and the indoor air capacitance $C$ (kWh/$^\circ$C). The input signals are the outdoor air temperature $T_\text{out}$ ($^\circ$C), the thermal power $\dot Q$ (kW) from the HP, and the thermal power $w$ (kW) from exogenous sources such as the sun, body heat, lights, plug loads, etc., as well as unmodeled dynamics. The 2R1C model, 
\begin{equation}
\begin{aligned}
C \frac{\text d T(t)}{\text d t} &= \frac{T_\text{out}(t) - T(t)}{R_\text{out}} + \frac{T_m(t) - T(t)}{R_m} \\
&+ \ \dot Q(t) + w(t) , \label{ct2R1C}
\end{aligned}
\end{equation}
can be reduced to an equivalent 1R1C model by defining an effective thermal resistance
\begin{equation}
R = \frac{R_m R_\text{out}}{R_m + R_\text{out}}
\label{effectiveResistance}
\end{equation}
and effective boundary temperature
\begin{equation}
\theta(t) = \frac{R_m T_\text{out}(t) + R_\text{out} T_m}{R_m + R_\text{out}}, 
\label{effectiveTemperature}
\end{equation}
giving the continuous-time dynamics
\begin{equation}
C \frac{\text d T(t)}{\text d t} = \frac{\theta(t) - T(t)}{R} + \dot Q(t) + w(t) . \label{ct1R1C}
\end{equation}
Assuming the input signals are piecewise constant over each time step of duration $\Delta t$ (h), the continuous-time dynamics discretize exactly to
\begin{equation}
\begin{aligned}
T(k+1) &= a T(k) + (1-a) \Big[ \theta(k) \\
& \quad + R \left( \dot Q(k) + w(k) \right) \Big] , \label{dt1R1C}
\end{aligned}
\end{equation}
where the integer $k$ indexes discrete time steps and
\begin{equation}
\label{eq:aDef}
a = \exp \left( - \frac{\Delta t}{R C} \right) .
\end{equation}

\subsubsection{System identification}
\label{system_identification}

Measured data from the house enable fitting of the unknown parameters $C$, $R_m$ and $R_\text{out}$, as well as the input signal $w(k)$. The model was fit to six weeks of data, with four weeks for training and two for validation. Identifying the thermal parameters required only about one week of data, while learning a predictive model of the exogenous thermal power $w(k)$ took about four weeks of data. The training data were gathered at five-minute intervals, consistent with the supervisory control time step. The fitting procedure first identifies the thermal circuit parameters, then trains a model to predict the exogenous thermal power (Section \ref{exogenous_section}). This section briefly describes the parameter estimation procedure; a more detailed description can be found in \cite{pergantis2024field}, while \cite{pergantisrcmodeling} compares this procedure to higher-order thermal circuit modeling approaches.

The parameter estimation procedure begins by selecting steady overnight periods, when d$T$/d$t \approx 0$, $T \approx T_m$, and $w \approx w_0$, a constant. During these periods, the continuous-time dynamics \eqref{ct2R1C} reduce to
\begin{equation}
T(t) - T_\text{out}(t) \approx \dot Q(t) R_\text{out} + c , \label{RoutFit}
\end{equation}
where $c = R_\text{out} w_0$. The unknown parameters $c$ and $R_\text{out}$ in Eq. \eqref{RoutFit} are fit to steady overnight $(T, T_\text{out}, \dot Q)$ data via linear regression. 

Given estimates of $R_\text{out}$ and $w_0 = c / R_\text{out}$, the remaining parameters $R_m$ and $a$ are fit to unsteady overnight periods, when the discrete-time dynamics \eqref{dt1R1C} reduce to
\begin{equation}
\begin{aligned}
&T(k+1) - \theta(k) - R \left( \dot Q(k) + w_0 \right) \\
\approx & \ a \left[ T(k) + \theta(k) + R \left( \dot Q(k) + w_0 \right) \right] . \label{aFit}
\end{aligned}
\end{equation}
The interior thermal mass temperature $T_m$ is assumed to be constant and approximated by the time-average indoor temperature. The fitting procedure combines a grid search over $R_m$ with linear regression. Each value of $R_m$, along with the estimated $R_\text{out}$, assume $T_m$, and measured $T_\text{out}$, defines corresponding values of the effective thermal resistance $R$ and effective boundary temperature $\theta(k)$ via Eqs. \eqref{effectiveResistance} and \eqref{effectiveTemperature}. This resolves all variables in Eq. \eqref{aFit} except the discrete-time dynamics parameter $a$, which is fit to unsteady overnight $(T, \theta, \dot Q)$ data for each $R_m$ value via linear regression. The final $(R_m, a)$ pair is selected to minimize the mean square error in approximating Eq. \eqref{aFit} in a validation dataset. Via Eq. \eqref{effectiveResistance}, the estimated $R_m$ and $R_\text{out}$ define the effective thermal resistance $R$, which is used for control along with the estimated $a$ and the discrete-time dynamics \eqref{dt1R1C}.

\subsection{Exogenous thermal power prediction}
\label{exogenous_section}

Current-limiting control requires accurate predictions of electricity demand over the next few time steps, especially the use of high-power resistance backup heat. Predicting the use of resistance backup heat requires accurately predicting the building's thermal load, and in particular the exogenous thermal power signal $w(k)$. This section develops an exogenous thermal power prediction method that combines an open-loop model (which does not use measurements of the building temperature state or past realizations of the exogenous thermal power as features) with a linear autoregressive model of the open-loop model's residual prediction errors. Candidate features for the open-loop model include functions of the time of day, day of week, and downloaded forecasts of weather variables.

In this approach, $\hat w(k + \ell | k)$, the full prediction made at time $k$ of the exogenous thermal power $w(k+\ell)$, can be written as
\begin{equation}
    \hat w(k + \ell | k) = \hat w^{\text{OL}}(k + \ell) + \hat e(k + \ell | k) ,
    \label{q_e_full}
\end{equation}
where $\hat w^{\text{OL}}(k + \ell)$ is the open-loop prediction and $\hat e(k + \ell | k)$ is the autoregressive error prediction. The training and validation dataset is formed by solving the discrete-time temperature dynamics  \eqref{dt1R1C} for $w(k)$, given estimates of $R$ and $a$ and  measurements of $T(k$), $\theta(k)$, $\dot Q(k)$, and $T(k+1)$:
\begin{equation}
w(k) = \frac{1}{R} \left( \frac{ T(k+1) - a T(k) }{ 1 - a } - \theta(k) \right) - \dot Q(k) . 
\label{qdote}
\end{equation}

Different model structures were investigated for the open-loop prediction $\hat w^\text{OL}$, including regression trees and random forests, Gaussian process regression, support vector machines (SVMs), and neural networks of varying depths. A range of predictive features were also investigated. After model selection and feature engineering, an SVM performed well with the following features: hour of day, outdoor temperature forecast, wind speed forecast, and global horizontal solar irradiance forecast. 

The left plot in Fig. \ref{qexog} shows the SVM predictions, $\hat w^{\text{OL}}(k + \ell)$, in training and validation data. The SVM captures the general trend well, with a validation root mean square prediction error of 2.4 kW. However, the SVM does not predict outliers well, particularly outliers due to the HP's defrost cycles. Further information on defrost is described in Section \ref{defrost_section} as part of the low-level controller. 

The relatively large residual open-loop prediction errors in Fig. \ref{qexog} motivated fitting $L$ autoregressive error prediction models of the form:

\begin{figure*}[tp]
    \centering
        \includegraphics[width=0.45\textwidth]{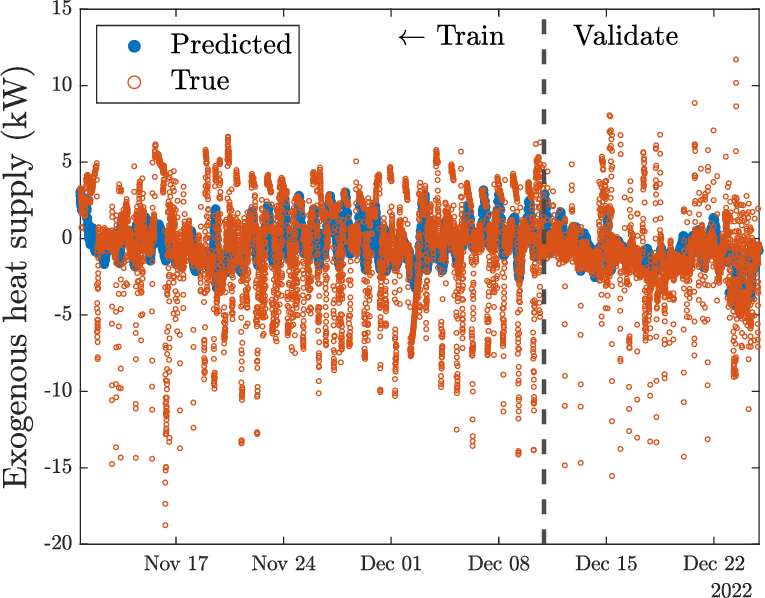}
        \qquad
        \includegraphics[width=0.46\textwidth]{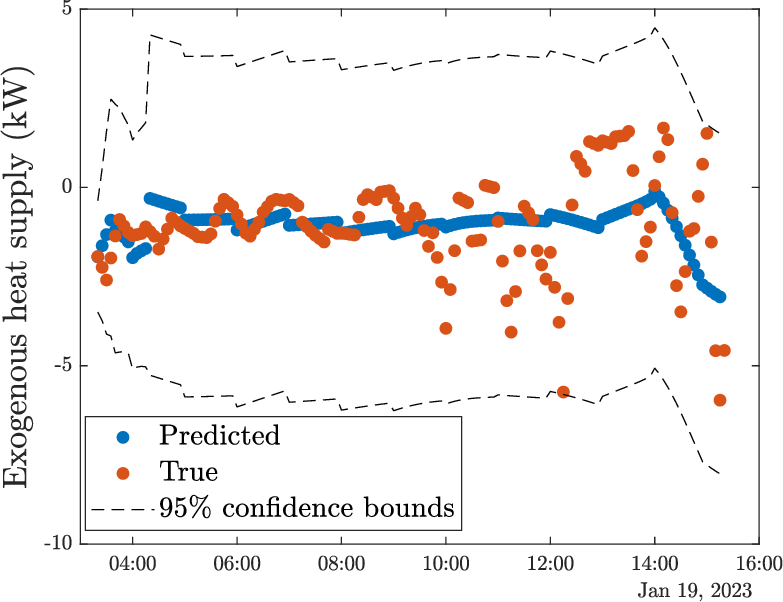}
        \caption{Left: Open-loop SVM predictions of the exogenous heat supply match validation data with a root mean square error of 2.4 kW, but miss outliers related to defrost cycles. Right: Full model predictions over 12 representative hours. The autoregressive error model improves predictions for the next hour, but adds little accuracy for longer-term forecasts.}
    \label{qexog}
\end{figure*}

\begin{equation}
\hat e(k + \ell | k) = \beta_{0,\ell} + \sum_{m=1}^{M_\ell} \beta_{m,\ell} e(k-m) 
\label{qdote_error_auto1}
\end{equation}
%Here $\hat{\dot Q}_e^\text{AR}$ is a constant baseline error, 
for $\ell = 0$, \dots, $L-1$. Here $\hat e(k + \ell | k)$ is the autoregressive prediction of $e(k + \ell)$ made at time $k$, $e(k-m) = w(k-m) - \hat w^\text{OL}(k-m)$ is the open-loop prediction error at time $k-m$ , and
\begin{equation}
\beta_{(\ell)} = \bmat 
\beta_{0,\ell} \\ 
\vdots \\ 
\beta_{M_\ell,\ell} \emat \in \R^{M_\ell+1}    
\end{equation}
is the parameter vector associated with the $\ell$-step-ahead autoregressive prediction. The memory $M_\ell$ of the $\ell$-step-ahead model is a hyperparameter that is tuned for each $\ell = 0$, \dots, $L-1$. The $\ell$-step-ahead model can be fit by forming a data set with feature matrix
\begin{equation}
X_{(\ell)} = \bmat
1& e(1) & \dots & e(M_\ell) \\
\vdots & \vdots & & \vdots \\
1& e(K-M_\ell) & \dots & e(K-1)
\emat
\label{eq:X_maxtrix}
\end{equation}
and target vector 
\begin{equation}
y_{(\ell)} = \bmat
e(M_\ell+\ell+1) \\
\vdots \\
e(K) \\
\emat .
\label{eq:target_vec}
\end{equation}
The model $y_{(\ell)} \approx X_{(\ell)} \beta_{(\ell)}$ can then be fit by least squares, giving the parameter estimate
\begin{equation}
    \hat \beta_{(\ell)} = (X_{(\ell)}^\top X_{(\ell)})^{-1} X_{(\ell)}^\top y_{(\ell)} .
\end{equation}

Hyperparameter tuning led to selecting an autoregressive prediction horizon of one hour (12 time steps). Beyond this horizon, the autoregressive model did not add much prediction accuracy. For this reason, the full prediction algorithm reverted to the open-loop prediction when predicting more than $L = 12$ steps ahead. 

The right-hand plot in Fig. \ref{qexog} shows predictions made by the open-loop SVM with autoregressive error correction over a 12-hour horizon beginning at 3:00 AM on a representative validation day. The dashed curves bound the 95\% Gaussian confidence intervals. The autoregressive error model improves short-term prediction accuracy, but beyond about one hour ahead, prediction accuracy degrades as the model reverts to the open-loop SVM predictions. 

\begin{figure*}[ht]
    \centering
        \includegraphics[width=0.45\textwidth]{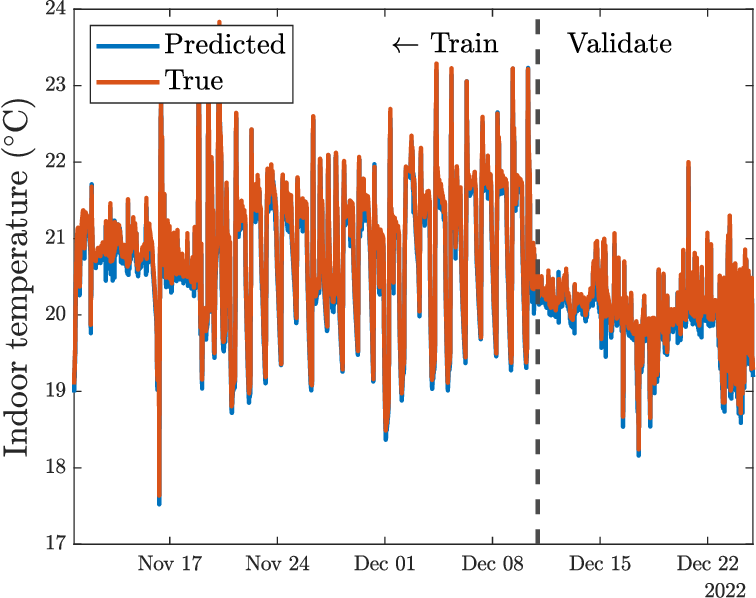}
        \qquad
        \includegraphics[width=0.45\textwidth]{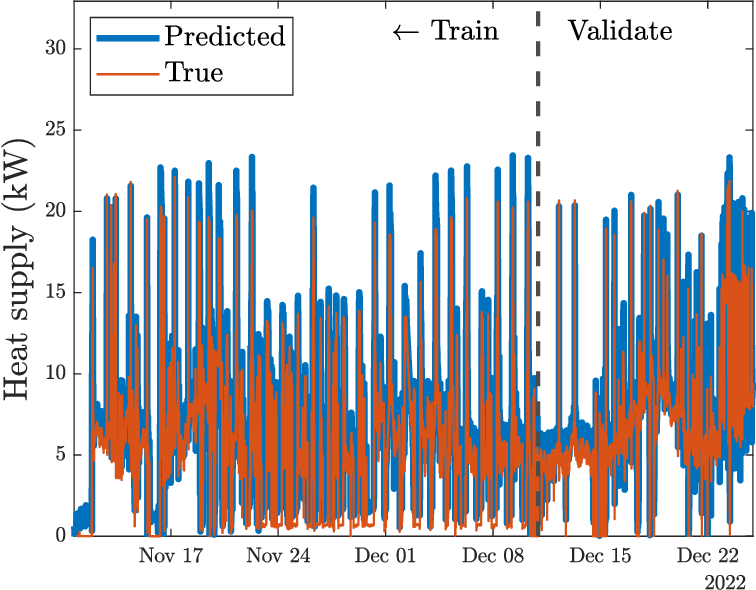}
        \caption{The full model's one-step-ahead predictions (blue curves) of the indoor temperature (left) and the HP thermal power (right) match the targets (orange curves) reasonably well in both training and validation. The validation root mean square temperature and power prediction errors were 0.2 $^\circ$C and 1.2 kW, respectively.}
    \label{modelFitFigure}
\end{figure*}

Fig. \ref{modelFitFigure} shows the one-step-ahead predictions made by the full model, including the thermal circuit model from Section \ref{rc_model} and the exogenous thermal power prediction models from Section \ref{exogenous_section}, in the training and validation sets. The validation root mean square errors were 0.2 $^\circ$C and 1.2 kW for predicting the indoor temperature $T$ and HP thermal power $\dot Q$, respectively. The HP thermal power predictions are made by substituting the predicted exogenous thermal power $\hat w(k+1 | k)$ from Eq. \eqref{q_e_full} and the measured temperatures $T(k)$ and $T(k+1)$ into the discrete-time dynamics \eqref{dt1R1C}, then solving for the HP thermal power $\dot Q(k)$.

\subsection{Equipment modeling}

\subsubsection{Heat pump}

\begin{figure}[ht!]
\centering
\includegraphics[width=0.45\textwidth]{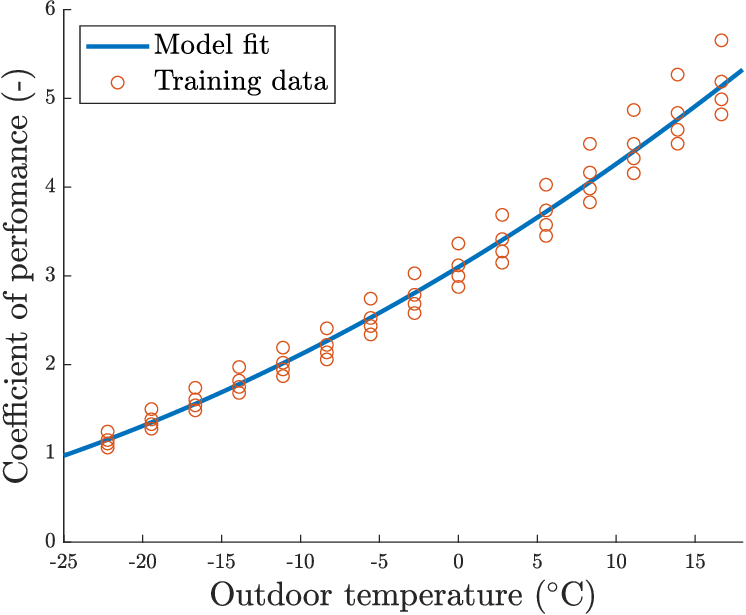}
\caption{Modeled only as a function only of the outdoor temperature, the HP COP matches training data with an $R^2$ of 0.98.}
    \label{COPFig}
\end{figure}

In principle, the COP $\eta$ of the test house's variable-speed air-to-air HP depends on the indoor temperature, outdoor temperature, and compressor speed. In practice, modeling the COP as a function only of the outdoor temperature gave sufficient accuracy for control purposes. Fig. \ref{COPFig} shows how a quadratic COP model fits data from the manufacturer for varying indoor and outdoor temperatures. The maximum rated electrical capacity of the compressor was modeled as $\overline P = 3.9$ kW and the minimum as $\underline P = 2.5$ kW. Both were modeled as constant due to limited data provided by the manufacturer. In this model, the HP can either be off or modulating in the interval $[\underline P, \overline P]$:
\begin{equation}
z(k) \underline P \leq \dot Q(k) / \eta(k) \leq z(k) \overline P ,
\end{equation}
where $z(k) \in \setof{0,1}$ indicates the HP's on/off state. The current draw from the HP is modeled as 
\begin{equation}
    I(k) = \frac{1}{V} \left[ \frac{\dot Q(k)}{\eta(k) f } + P_r(k) \right] ,
    \label{current_hp}
\end{equation}
where $V = 240$ V is the rated voltage at which the HP and backup heating element draw power and $f = 0.8$ is the HP power factor (assumed constant). The power factor of the backup heating element, a purely resistive load, is assumed to be unity. The backup heating element power is either zero or in one of three discrete stages: $P_r(k) \in \setof{0, 9.6, 14.4, 19.2}$ kW.

\subsubsection{Water heater}
\label{WH_model}

The WH in the test house is equipped with one flow meter and three thermocouples. The flow meter was installed on the outlet pipe. Thermocouples were placed on the upper and lower parts of the tank, both adjacent to the manufacturer's thermistors, and at the tank's inlet water line. Measurements are read by a central data acquisition system, then communicated to a cloud database where all data from the house are stored. From observing the operation of the tank in resistance-only mode, only the bottom heating element is initiated and the top element does not turn on, resulting in a 4.5 kW capacity in electric mode assuming a conversion efficiency of 100\% from electrical to thermal power. 

The water temperature $T_w$ ($^\circ$C) was assumed to be spatially uniform throughout the tank and modeled as a weighted average of the readings of the top thermistor (75\% weight) and bottom thermistor (25\% weight), consistent with observations from \cite{osti_1140094, 9087716}. This is a reasonable model for WH in resistance mode, as the two thermocouples typically read within 2$^{\circ}$C of one another except during water draws. During water draws, cold water moves into the bottom of the tank, balancing hot water drawn from the top of the tank. The inlet water pushes upward the water heated by the bottom resistor. For these reasons, the bottom thermocouple reading approaches the inlet water temperature during water draws, while the top thermocouple reading remains nearly constant. After water draws, the tank temperature gradually returns to an approximately spatially uniform state.

A first-order linear thermal circuit, similar to the building envelope model described in Section \ref{rc_model}, models the water temperature dynamics:
\begin{equation}
    \label{eq:discretized water}
    \begin{aligned}
        T_w(k+1) = &\ a_w T_w(k) + (1 - a_w) \Big[ T_a \\
        &\qquad + R_w \Big( P_w(k) - w_w(k) \Big) \Big] ,
    \end{aligned}
\end{equation}
where
\begin{equation}
    a_w = \exp \left( - \frac{\Delta t}{R_w C_{w}} \right) .
\end{equation}
In this model, the thermal capacitance $C_w = 0.197$ (kWh/$^\circ$C) is the product of the mass of stored water and the specific heat of water at 50 $^\circ$C. The thermal resistance $R_w = 1476$ ($^\circ$C/kW) is the reciprocal of the product of the surface area and thermal transmittance of the tank wall. The thermal resistance determines the rate of heat transfer from the water to the surrounding air, which is assumed to have constant temperature $T_a = 19$ $^\circ$C. The heat transfer rate out of the WH due to water draws, $w_w$ (kW), is the product of the volumetric flow rate measured by the flow meter, the density of water, the specific heat of water, and the difference between the temperatures measured by the top thermistor and the inlet water thermistor. The rate of heat transfer to the tank from the heating elements equals the measured input electric power, $P_w$ (kW). The WH is assumed to have unity power factor and to perfectly track water temperature set-points.

\begin{figure}
\centering
\includegraphics[width=0.45\textwidth]{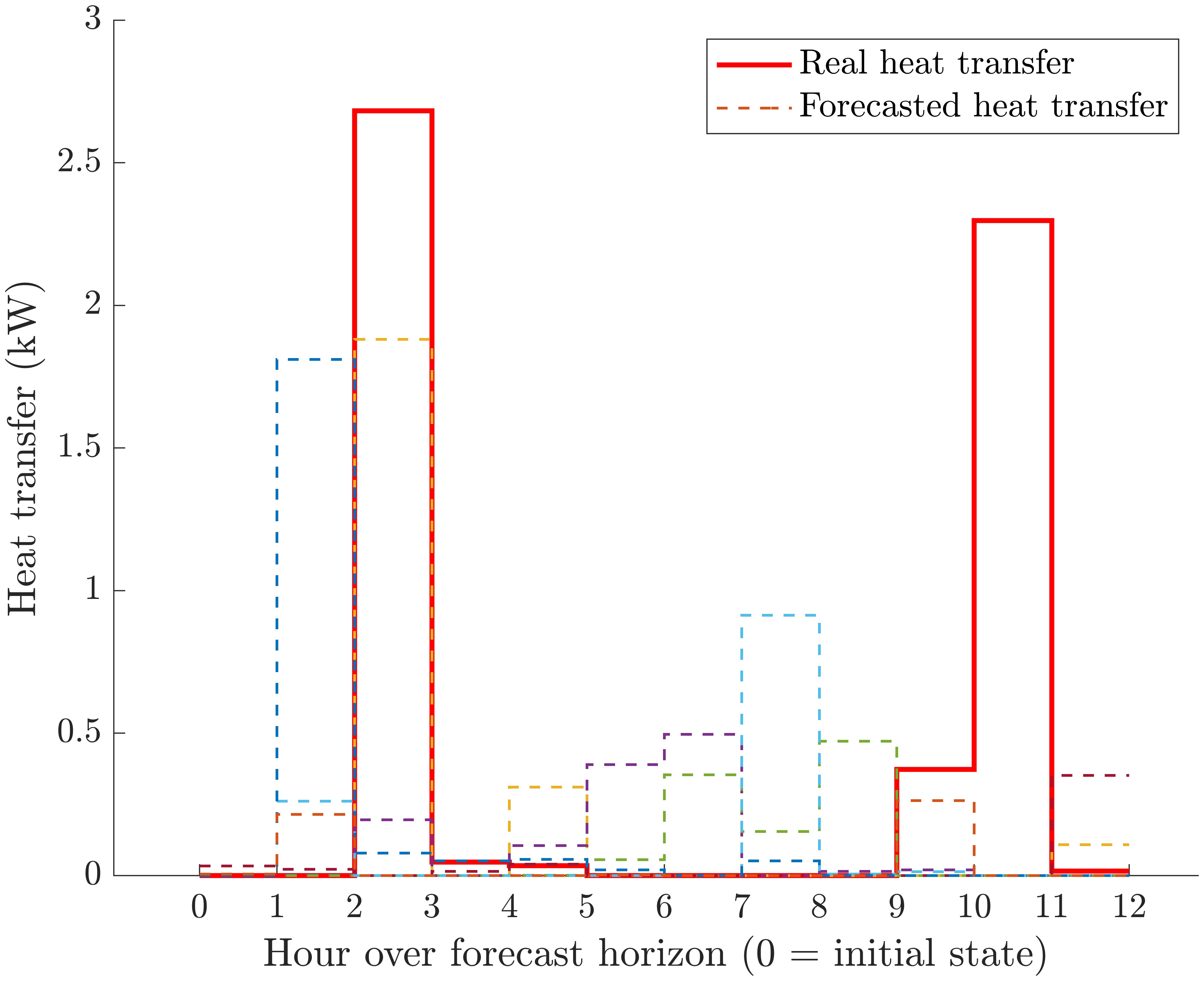}
\caption{Real (solid red curve) and forecasted (dashed curves) heat transfer from hot water draws on a typical day.}
\label{Water Forecast}
\end{figure}

\begin{figure*}
    \centering
        \includegraphics[width=0.45\textwidth]{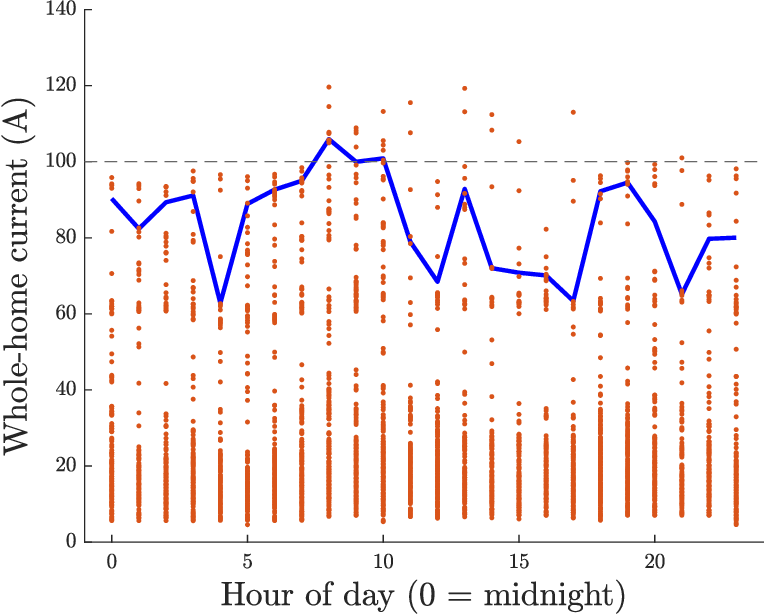}
        \qquad
        \includegraphics[width=0.45\textwidth]{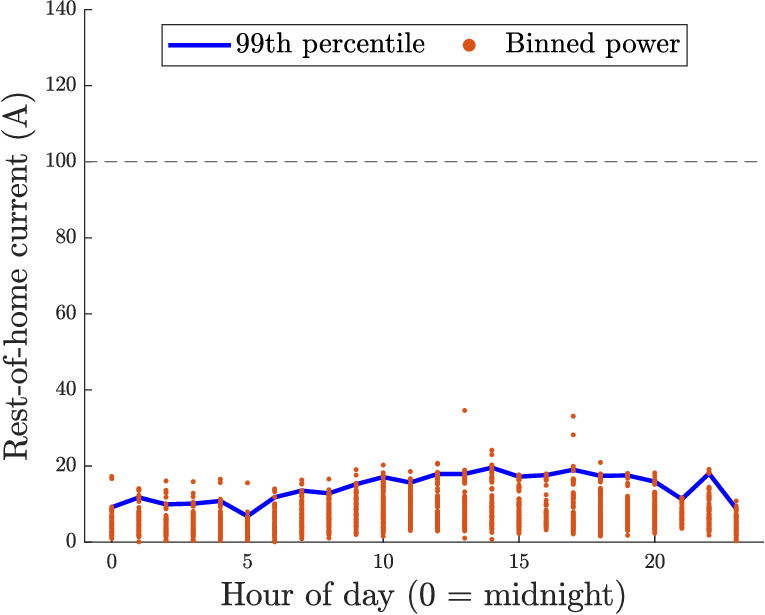}
        \caption{Left: Whole-house current over 43 days from the winter of 2022-23, binned by hour of day. Right: Current from all loads other than the controlled HP and WH.}
    \label{binned_current_draw}
\end{figure*}

A nearest neighbors algorithm was developed to forecast the heat transfer from hot water draws, $w_w$. The authors chose the nearest neighbors algorithm because several other methods for time-series forecasting did not preform well. The water draw signal was challenging to predict, as water draws are often zero or nearly zero, but spike during baths, showers, and to a lesser extent during cycles of the dishwasher and clothes washing machine \cite{Premer_HPWH}. The number, timing, and magnitudes of these spikes vary significantly from day to day and do not show a simple autocorrelation structure. The nearest neighbors algorithm generates multiple forecast scenarios by resampling historical water draw data. Using these multiple forecasts in supervisory control improved robustness to large water draws at unusual times.

At each time step, the nearest neighbors algorithm compares the previous eight hours of heat transfer data from water draws, recorded hourly, with the last thirty days of data. The closest matching days are identified using the mean absolute error. Historical heat transfer profiles from these nearest days become forecasts of the next 12 hours. Fig. \ref{Water Forecast} shows the actual heat transfer (solid red curve) and four forecasts (dashed curves) on a typical day. While two forecasts nearly capture the large draw that happens two hours ahead, none of the forecasts predict the second large draw that happens ten hours ahead. This exemplifies a typical pattern of higher accuracy for shorter-term predictions. For this reason, the nearest neighbors algorithm runs at each supervisory control time step, regularly generating new forecasts that account for recent water usage. To increase robustness to prediction errors, the supervisory controller uses a tunable number $S$ of nearest neighbors as forecast scenarios. More information on this methodology is presented in \cite{Premer_HPWH}.

\subsubsection{Uncontrolled loads}
\label{rest_of_house}

In addition to the controlled HP and WH, the test house has a variety of other electrical loads. Some of the other loads, such as the plug-in hybrid EV, dryer, and dishwasher, are potentially controllable devices with load-shifting capabilities \cite{SALPAKARI2016425}. However, controlling these devices would require adding sensing and communicating infrastructure and extending control algorithms. Therefore, this paper controls only the HP and WH, leaving occupants with complete control over all other devices. The combined power used by all other devices is modeled as a single disturbance term.

Accurately predicting the uncontrolled electrical load proved challenging, as the graduate student occupants of the test house do not follow typical workday routines. Therefore, rather than attempting to predict the exact uncontrolled load, historical data were used to compute 99\% confidence intervals for each hour of the day. The supervisory controller was then designed to be robust to these (nearly) worst-case realizations of the uncontrolled load. Fig. \ref{binned_current_draw} shows historical data and confidence intervals for the whole-house current (left plot) and the current from the uncontrolled loads (right plot). These plots show that the current from uncontrolled loads ($\simm$10-20 A at the 99th percentile) is a relatively small part of the whole-house current ($\simm$70--110 A). The controlled HP and WH make up the difference between the left and right plots.

\subsection{MPC problem formulation}

This section formulates the optimization problem that the supervisory MPC system solves at each time step. The goals are to reduce energy costs, user discomfort from low space or water temperatures, and the frequency and magnitude of whole-home current limit violations. There is significant uncertainty from unmodeled dynamics, imperfect weather forecasts, unexpected occupant behavior, and other factors. Any of these uncertainties could cause discrepancies between the true and modeled consequences of control actions, jeopardizing occupant comfort or putting electrical infrastructure at risk. To mitigate these risks, the MPC system handles uncertainty in four ways. First, it hedges against multiple scenarios of future water draws, which are difficult to predict. Second, it requires a high probability of satisfying current constraints. Third, it uses pessimistic forecasts of the exogenous thermal power from sunlight, bodies, plug loads, unmodeled dynamics, etc. Fourth, the MPC system uses feedback to compensate for model mismatch and forecast errors, solving a new optimization problem at each time step with updated measurements and forecasts. 

To avoid burdensome notation, this section describes only the MPC problem solved at time zero. At each subsequent time step, the system evolves to a new state, new measurements and forecasts are obtained, and an updated version of the same problem is solved. Eqs. \eqref{eq:objective}-\eqref{eq:integer_variables} show the full problem formulation. These equations use the set definitions $\mathcal K = \setof{0, \dots, K-1}$ and $\mathcal S = \setof{1, \dots, S}$.

\begin{figure*}[t]
\centering
\begin{subequations}
\label{eq:full_formulation}
\begin{align}
&\text{\hspace{2em} Objective:} &&\text{minimize } \pi_I \max_{s \in \mathcal S} \max_{k \in \mathcal K} \max\setof{ 0, I^s(k) + I_w^s(k) + q_\alpha(k) - \overline I }  \nonumber \\
&&& \quad + \frac{ \Delta t }{S} \sum_{s \in \mathcal S} \sum_{k \in \mathcal K} \Bigg[ \pi_e \left( \frac{\dot Q^s(k)}{\eta(k)} + \frac{z_r^s(k) \overline P_r}{4} + z_w^s(k) \overline P_w \right) \nonumber \\ 
&&& \quad + \pi_t(k) | \tilde T^s(k) - T^\dagger(k) | + \pi_w(k) | T_w^s(k) - T^\dagger_w(k) | \Bigg ] \label{eq:objective} \\
&\text{\hspace{2em} Indoor temperatures:} &&T^s(0) = T_0 , \ T^s(k+1) = a T^s(k) + (1 - a) \bigg[ \theta(k) + R \bigg( \dot Q^s(k) \nonumber \\ 
&&& \quad + \frac{z_r^s(k) \overline P_r}{4} + \hat w(k | 0) - \lambda(k) \sigma(k) \bigg) \bigg]  \ \forall \ k \in \mathcal K, \ s \in \mathcal S \label{eq:state_air} \\
&\text{\hspace{2em} Set-point tracking:} &&T^s(k+1) = b T^s(k) + (1-b) (\tilde T^s(k+1) - \gamma)  \ \forall \ k \in \mathcal K, \ s \in \mathcal S \label{eq:setpoint_tracking} \\
&\text{\hspace{2em} Water temperatures:} &&T_w^s(0) = T_{w0}, \ T_w^s(k+1) = a_w T_w^s(k) + (1 - a_w) \big[ T_a \nonumber \\
&&& \quad + R_w \left( z_w^s(k) \overline P_w - w_w^s(k)  \right) \big]  \ \forall \ k \in \mathcal K, \ s \in \mathcal S \label{eq:state_water} \\
&\text{\hspace{2em} HP and WH currents:}
&&I^s(k) =\frac{1}{V} \left[ \frac{\dot Q^s(k)}{ \eta(k) f}  + \frac{ z_r^s(k) \overline P_r}{4} \right]  \ \forall \ k \in \mathcal K, \ s \in \mathcal S  \label{eq:HP_current} \\ 
&&&I_w^s(k) = z_w^s(k) \overline P_w / V \label{eq:WH_current}  \ \forall \ k \in \mathcal K, \ s \in \mathcal S \\
&\text{\hspace{2em} Temperature bounds:} &&\underline T \leq T^s(k+1) \leq \overline T  \ \forall \ k \in \mathcal K, \ s \in \mathcal S \label{eq:air_comfort} \\
&&&\underline T_w \leq T_w^s(k+1) \leq \overline T_w  \ \forall \ k \in \mathcal K, \ s \in \mathcal S \label{eq:water_comfort} \\
&\text{\hspace{2em} HP capacity:} &&z^s(k) \underline P \leq \dot Q^s(k) / \eta(k) \leq z^s(k) \overline P  \ \forall \ k \in \mathcal K, \ s \in \mathcal S \label{eq:HP_capacity} \\
&\text{\hspace{2em} Action consistency:} && \tilde T^s(0) = \tilde T^1(0), \ T_w^s(1) = T_w^1(1) \ \forall \ s \in \setof{2, \dots, S} \label{eq:action_consistency} \\
&\text{\hspace{2em} Integer variables:} 
&& z^s(k) \in \{ 0,1 \}, \ z_r^s(k) \in \{ 0,2,3,4 \}, \ z_w^s(k) \in \{ 0,1 \}  \ \forall \ k \in \mathcal K, \ s \in \mathcal S
\label{eq:integer_variables}
\end{align}
\end{subequations}
\end{figure*}

\subsubsection{Water draw scenarios}
\label{waterScenarios}

As discussed in Section \ref{WH_model}, water draws are difficult to predict. To add robustness to prediction errors, the MPC formulation uses $S$ scenarios of the future water draw trajectory, rather than a single point forecast. The scenarios $w_w^1(k)$, \dots, $w_w^S(k)$ are the $S$ nearest neighbors in the water draw forecasting algorithm. Fig. \ref{Water Forecast} shows four scenarios on a typical day. The experiments reported in Section \ref{testing_results} used $S = 8$ scenarios, a number that struck a reasonable balance between computation time and robustness to prediction errors. Associated with the $S$ water draw scenarios are $S$ trajectories of each state and action variable. In Eq. \eqref{eq:state_water}, for example, each water temperature trajectory $T_w^s(0)$, \dots, $T_w^s(K)$ is initialized at $T_w^s(0) = T_{w0}$ and satisfies the water temperature dynamics \eqref{eq:discretized water} with $w_w(k)$ replaced by $w_w^s(k)$.

\subsubsection{Decision variables}

The decision variables include the indoor temperature scenarios $T^s(k)$ and water temperature scenarios $T_w^s(k)$, all defined for $k = 0$, \dots, $K$ and $s  \in \mathcal S$. The decision variables also include the following, defined for $k = 0$, \dots, $K-1$ and $s  \in \mathcal S$: The indoor temperature set-point $\tilde T^s(k)$, HP thermal power output $\dot Q^s(k)$, HP input current $I^s(k)$, WH input current $I_w^s(k)$, HP on/off state $z^s(k) \in \setof{0,1}$, WH on/off state $z_w^s(k) \in \setof{0,1}$, and a variable $z_r^s(k) \in \setof{0, 2, 3, 4}$ defined such that the resistance heat can turn off or occupy any of three non-zero stages. The resistance heat power satisfies $P_r^s(k) = z_r^s(k) \overline P_r / 4 \in \setof{0, 9.6, 14.4, 19.2}$ kW when $\overline P_r = 19.2$ kW, consistent with the staging behavior described in Section \ref{testHouse}.

\subsubsection{Scenarios and control actions}

The control actions at time $k = 0$ in scenario $s$ are the indoor temperature set-point $\tilde T^s(0)$ and the water temperature set-point $\tilde T_w^s(0)$. While the decision variables include separate control action trajectories for each of the $S$ scenarios, the control system must send a unique, unambiguous set-point to the HP, and similarly for the WH. To ensure that the control actions at time $k = 0$ are the same across all scenarios, the optimization enforces the `action consistency' constraints $\tilde T^1(0) = \dots = \tilde T^S(0)$ and $T_w^1(1) = \dots = T_w^S(1)$ in Eq. \eqref{eq:action_consistency}. (Under the assumption of perfect set-point tracking by the device-level WH controller, the first water temperature set-point, $\tilde T_w^s(0)$, equals the next planned water temperature, $T_w^s(1)$.) The `action consistency' constraints apply only at the first time step; at all subsequent time steps, a separate action plan can be tailored to each scenario. After solving the optimization problem at time $k = 0$, the high-level controller sends $\tilde T^1(0)$ and $T_w^1(1)$ to the low-level controller, the system evolves to a new state, new measurements and forecasts are obtained, and a new optimization problem is solved to generate the control actions at time $k = 1$.

\subsubsection{Input data}

The time-invariant input data are the current limit violation price $\pi_I = 10^6$ \$/A, current limit $\overline I = 100$ A, time step $\Delta t = 0.083$ h (five minutes), number of water draw scenarios $S = 8$, number of time steps $K = 144$ (12-hour horizon), electricity price $\pi_e = 0.14$ \$/kWh, backup resistance power capacity $\overline P_r = 19.2$ kW, WH power capacity $\overline P_w = 4.5$ kW, space and water temperature dynamics parameters $a = 0.9$ and $a_w = 0.9997$, space and water thermal resistances $R = 0.9$ $^\circ$C/kW and $R_w = 1476$ $^\circ$C/kW, indoor temperature tracking dynamics parameter $b = 0.5$, space thermostat dead-band half-width $\gamma = 0.5$ $^\circ$C, ambient temperature $T_a = 19$ $^\circ$C around the water tank, initial space and water temperatures $T_0$ and $T_{0w}$ (both measured at each time step), voltage $V = 240$ V, HP power factor $f = 0.8$, indoor temperature bounds $\underline T = 18.5$ $^\circ$C and $\overline T = 23$ $^\circ$C, water temperature bounds $\underline T_w = 43.3$ $^\circ$C and $\overline T_w = 60$ $^\circ$C, and HP electric power bounds $\underline P = 2.5$ kW and $\overline P = 3.9$ kW.

The time-varying input data are the $\alpha$-quantile $q_\alpha(k)$ of the uncontrolled current $I_u(k)$ from all devices other than the HP and WH (discussed in Section \ref{currentLimit}), indoor temperature preference $T^\dagger(k)$ (21 $^\circ$C during the day and 19.5 $^\circ$C overnight), water temperature preference $T^\dagger_w(k)$ (\S\ref{priceTuning}), space heating discomfort price $\pi_t(k)$ (\S\ref{priceTuning}), water heating discomfort price $\pi_w(k)$ (\S\ref{priceTuning}), HP COP $\eta(k)$ (\S\ref{weatherData}), effective boundary temperature $\theta(k)$ (\S\ref{weatherData}), exogenous thermal power prediction $\hat w(k|0)$ (\S\ref{safetyMargins}), $k$-step-ahead prediction standard deviation $\sigma(k)$ (\S\ref{safetyMargins}), tunable pessimism parameter $\lambda(k)$ (\S\ref{safetyMargins}), and scenarios $w_w^1(k)$, \dots, $w_w^S(k)$ of the thermal power from water draws (\S\ref{waterScenarios}).

\subsubsection{Objectives}

The objectives include reducing the energy cost
\begin{equation}
\pi_e \left( \frac{\dot Q^s(k)}{\eta(k)} + \frac{z_r^s(k) \overline P_r}{4} + z_w^s(k) \overline P_w \right) 
\end{equation}
and the costs of discomfort due to indoor temperatures, 
\begin{equation}
\pi_t(k) | \tilde T^s(k) - T^\dagger(k) | ,
\end{equation}
and to water temperatures,
\begin{equation}
\pi_w(k) | T_w^s(k) - T^\dagger_w(k) |. 
\end{equation}
These costs are all integrated over time and averaged over scenarios. The objectives also include reducing a penalty on the whole-home current, described below.

\subsubsection{Whole-home current limit}
\label{currentLimit}

The whole-home current in each scenario $s$ is the sum of the HP and backup resistance heater current $I^s(k)$ (a decision variable), the WH current $I_w^s(k)$ (a decision variable), and the current $I_u(k)$ drawn by uncontrolled loads (a random variable). Ideally, the whole-home current should remain below the deterministic limit $\overline I = 100$ A with a high probability $\alpha$:
\begin{equation}
\mathop{\bf Prob} \setof{ I^s(k) + I_w^s(k) + I_u(k) \leq \overline I } \geq \alpha . \label{currentConstraint}
\end{equation}
Here the probability refers to the distribution of $I_u(k)$, the only random variable in Eq. \eqref{currentConstraint}. This holds if
\begin{equation}
I^s(k) + I_w^s(k) + q_\alpha(k) \leq \overline I ,
\end{equation}
where $q_\alpha(k)$ is the $\alpha$-quantile of $I_u(k)$. The right-hand plot of Fig. \ref{binned_current_draw} shows empirical estimates of $q_\alpha(k)$ with $\alpha = 0.99$.

In practice, enforcing these constraints often led to infeasible MPC problems. Furthermore, as discussed in Section \ref{problemStructure}, rare violations of the current limit may be acceptable, but the violations must not be too large or the main circuit breaker may trip. For these reasons, the MPC problem treats the whole-home current limit as a soft constraint, penalizing the violation magnitude
\begin{equation}
\max \setof{ 0, I^s(k) + I_w^s(k) + q_\alpha(k) - \overline I} 
\end{equation}
in the worst case (over time and scenarios) at a tunable price $\pi_I$.

\subsubsection{Space heating safety margins}
\label{safetyMargins}

Over-predicting the thermal power from exogenous heat sources (such as bodies, lights, plug loads and the sun) can risk violating the whole-home current limit. If the true exogenous thermal power $w(k)$ is less than the prediction $\hat w(k | 0)$ made at time zero, then the building will require more heat at time $k$ than anticipated and the HP will draw more current than anticipated. 

For this reason, the MPC formulation pessimistically assumes that $w(k)$ will be less than the prediction $\hat w(k | 0)$. Mathematically, $\hat w(k | 0)$ is replaced by 
\begin{equation}
\hat w(k | 0) - \lambda(k) \sigma(k) ,
\end{equation}
where $\sigma(k)$ is the standard deviation of the $k$-step-ahead autoregressive error model in the validation data. The tunable parameter $\lambda(k)$, which governs the degree of pessimism in forecasting $w(k)$, was tuned to 1 when the outdoor temperature at time $k$ was below -10$^\circ$C, and to 1.5 otherwise.

\subsubsection{Indoor temperature set-point tracking}
\label{tracking}

Under baseline operations, the manufacturer's controls on the variable-speed HP imperfectly track indoor temperature set-point changes. Set-point tracking imperfections include both delays and a steady-state offset, as the manufacturer controls drive the measured indoor temperature to the low end of the thermostat dead-band rather than to the set-point itself. The closed-loop tracking dynamics approximately satisfy the first-order model
\begin{equation}
\frac{\text d T(t)}{\text d t} = \frac{ \tilde T(t) - \gamma - T(t) }{\tau} ,
\end{equation}
where $\tau$ (h) is the tracking time constant and $\gamma = 0.5$ $^\circ$C is the indoor thermostat dead-band half-width. Assuming the set-point $\tilde T(t)$ is piecewise constant, the tracking model discretizes exactly to
\begin{equation}
T(k+1) = b T(k) + (1-b) (\tilde T(k) - \gamma) ,
\end{equation}
where
\begin{equation}
b = \exp \left( - \frac{\Delta t}{\tau} \right) .
\end{equation}
The time constant $\tau$ was calibrated by observing the responses to several step changes to the indoor temperature setpoint $\tilde T$. After a step change to $\tilde T$, the manufacturer controls typically eliminated 95\% of the initial error between the measurement $T$ and the reference $\tilde T - \gamma$ after 20 to 40 minutes. (A first-order linear system eliminates 95\% of the initial tracking error after three time constants.) Based on step-response observations and simulation experiments, the parameter $b$ was manually tuned to 0.5, corresponding to a time constant of $\tau = -5/\ln(0.5) = 7.2$ minutes. In the authors' experience, this is a fairly typical time constant for indoor temperature set-point tracking from a variable-speed air-to-air heat pump manufacturer's device-level control system. The tracking time constant $\tau$ is distinct from the time constant $RC$ associated with the passive response of the indoor temperature to changing weather conditions. For single-family houses, the time constant $RC$ is typically on the order of several hours.

%This led to setting $\tau = 7$ minutes (a first-order linear system eliminates 95\% of the initial tracking error after three time constants) and $b = \exp(-5/7) = 0.49$.

\subsubsection{Price tuning}
\label{priceTuning}

The objective function in Eq. \eqref{eq:objective} balances the competing goals of reducing (1) the worst-case whole-home current limit violation magnitude, (2) the total energy use, and (3-4) deviations of the indoor temperature and water temperature from the occupants' preferences. Trade-offs between these competing goals are governed by the prices $\pi_I$ of current violations, $\pi_e$ of electricity, $\pi_t(k)$ of indoor temperature deviations, and $\pi_w(k)$ of water temperature deviations. The electricity price $\pi_e = 0.14$ \$/kWh was obtained from the test house's utility rate plan. The price $\pi_I$ was manually tuned to 10$^6$ \$/A to strongly penalize current violations. 

For the indoor temperature deviation price $\pi_t(k)$, previous work \cite{pergantis2024field} used simulations at each time step to adaptively balance energy costs against a comfort metric called the Predicted Percentage Dissatisfied \cite{tartarini2020pythermalcomfort}. Due to computation time constraints, this paper uses a simple linear model fit to simulation results from a range of weather conditions:
\begin{equation}
    \pi_t(k) = -0.24 T_\text{out}(k) + 1.2 ,
\end{equation}
in units of \$$/^\circ$C/h. This price structure imposes higher penalties on indoor temperature deviations when it is colder outside. 

The water temperature deviation price $\pi_w(k)$ was initially set to zero. This resulted in low water set-points most of the time, with preheating only before forecasted water draws. This behavior led to unsatisfactory comfort during cold weather. To improve comfort, two preheating windows were defined: 4 AM to 6 AM and noon to 2 PM. During preheating windows, the price $\pi_w(k)$ was 12 \$$/^\circ$C/h and the preference $T_w^\dagger(k)$ was 130 $^\circ$F (54.4 $^\circ$C). At all other times, the control system used $\pi_w(k) = 6$ \$/$^\circ$C/h and $T_w^\dagger(k) = 120$ $^\circ$F (48.9 $^\circ$C). Section \ref{occupantComfort} discusses comfort and water temperatures in more detail.

\subsubsection{Weather-dependent data}
\label{weatherData}

The control system pulls weather data and forecasts from the Oikolab weather service \cite{oikolab} via an Application Programming Interface. The COP $\eta(k)$ is computed by propagating the outdoor temperature forecast through the quadratic model in Fig. \ref{COPFig}. As discussed in Section \ref{rc_model}, the effective boundary temperature $\theta(k)$ is a weighted average of the outdoor temperature forecast and the thermal mass temperature $T_m$. As discussed in Section \ref{exogenous_section}, central predictions of the exogenous thermal power, $\hat w(k | 0)$, are computed by propagating weather forecasts and time-of-day information through the trained SVM, then propagating historical measurements through the autoregressive models to predict the error in the SVM forecast.

\subsubsection{Implementation details}

The MPC problem solved at each time step is a mixed-integer convex program. The supervisory control system solves this problem in Matlab using the CVX optimization toolbox \cite{cvx}. CVX introduces ancillary variables to transform the MPC problem, which as specified in Eqs. \eqref{eq:objective}-\eqref{eq:integer_variables} is nonlinear, into a standard-form mixed-integer linear program. CVX passes the transformed problem to the Gurobi 9 solver, then inverse-transforms the solution back to the form of Eqs. \eqref{eq:objective}-\eqref{eq:integer_variables}. The transformed problem solved by Gurobi has 19,180 decision variables and 12,122 constraints. The full computation process, including transformations and numerical solution to an optimality gap of less than 1\%, typically takes at most two minutes with a 2.6 GHz i7-9750H processor and eight GB of memory. This leaves about three minutes within each five-minute time step for the low-level control logic and for communication with sensors and actuators. Once the problem is solved, the high-level controller sends the resulting HP and WH thermostat set-points to the low-level controller, which may push them to the thermostats or override them based on real-time measurements of house conditions (see Section \ref{low_level_controller}). About three minutes after each MPC solution finishes, new measurements and forecasts arrive and the next MPC solution begins. Thus, absent overrides by the low-level controller, each MPC action is implemented for five minutes. Fig. \ref{flowchart} illustrates information flow between the high- and low-level controllers. Section \ref{future_work} discusses ways to reduce the time required for computation.

\section{Low-level controller}
\label{low_level_controller}

The high-level controller operates at a five-minute time step, aiming to maintain comfortable air and water temperatures, reduce energy costs, and maintain the whole-home current below a safe limit. The high-level controller accomplishes this in part by shifting space- and water-heating load using disturbance forecasts and numerical optimization. However, unanticipated disturbances within a five-minute time step, such as HP defrost cycles, can still put electrical infrastructure at risk. This motivated developing a low-level controller with a time step of 30 seconds. The low-level controller quickly reacts to unanticipated current spikes by lowering the set-points of the WH or HP.

\subsection{Control logic}

\begin{figure}[ht!]
    \centering   \includegraphics[width=0.47\textwidth]{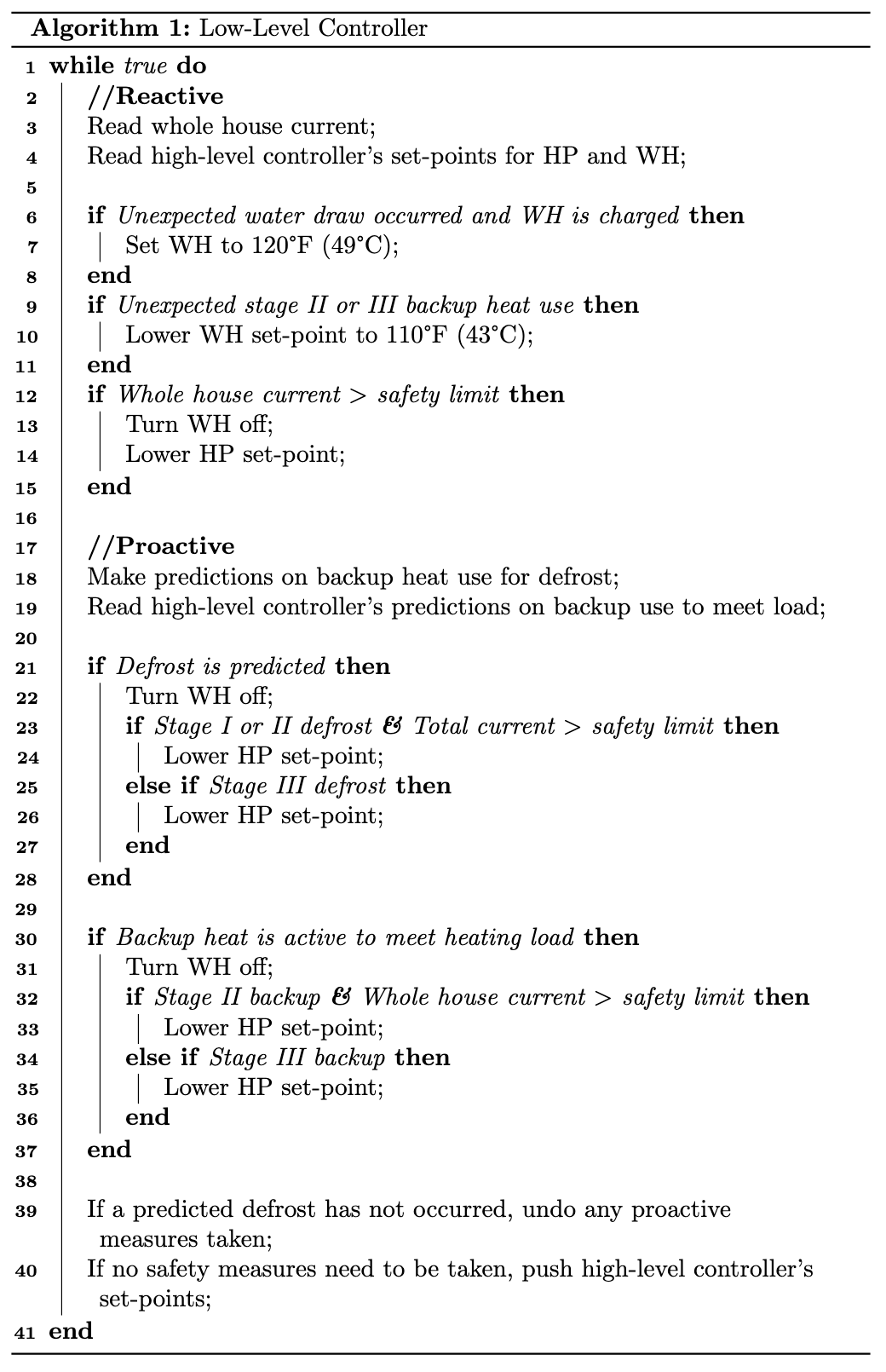}
    \caption{Pseudocode of the low-level control logic, which continuously monitors both the outputs of the high-level controller and the real-time house conditions.}
    \label{fast_layer_pseudo}
\end{figure}

The low-level controller can: (a) predict if the HP will request a defrost cycle by monitoring the real-time supply air temperature (Section \ref{defrost_section}), and (b) receive predictions from the high-level controller if backup heat is predicted over the next five minutes, and if so, which stage of backup heat. Based on these predictions, as well as real-time measurements of the whole-home current, the low-level controller adjusts the operation of the HP and WH as shown in Fig. \ref{fast_layer_pseudo}.

The low-level controller continuously monitors the indoor conditions and the decisions of the high-level controller, iterating every 30 seconds. The reactive portion starts by reading the real-time whole-home current, the backup heat status, and water usage in the house, with which the following three reactive (if-else) tests are performed. 

\begin{enumerate}
    \item If there is significant water use and the WH is sufficiently charged (meaning its temperature is above 120 \textdegree F or 49 \textdegree C), the WH set-point is lowered to 120 \textdegree F (49 \textdegree C). This is an energy efficiency measure.
    
    \item If unexpected use of Stage II or III resistance heat is detected, turn off the WH. Stages II and III backup heat are rated at 60 A and 80 A, respectively, leaving limited capacity for additional current draws. For comfort purposes, the two WH charging periods (scheduled during low-backup usage times) result in sufficient comfort. %(Sections \ref{weatherData}, \ref{testing_results}).
    
    \item If the whole-home current is above the safety limit, all possible measures are taken to lower the current: turning off the WH and lowering the HP thermostat set-point.
\end{enumerate}

For the proactive features, the low-level controller first predicts potential defrost cycles occurring in the next time step through an autoregressive model on the historical HP supply temperatures (as described in Section \ref{defrost_section}). If defrost is predicted in the next time step, then the WH is turned off immediately. If Stage I or II defrost is predicted, then the HP set-point is lowered only if the net current draw is above the safety limits. If Stage III defrost is predicted, then the HP set-point is lowered regardless. This is in place to avoid a fast ramp-up of the compressor when backup heat is on and the HP switches to heating after the completion of the defrost cycle. The proposed methodology prioritizes the HP to ensure satisfactory indoor temperatures. This can result in undercharging the WH unless care is taken to pre-charge the WH before times of high water draws (see Sections \ref{priceTuning} and \ref{occupantComfort}).

\subsection{Defrost prediction}
\label{defrost_section}

Air-source HPs periodically run defrost cycles to maintain heating capacity. In heating mode, air-source HPs absorb energy from the outdoor air by driving very cold refrigerant through the outdoor heat exchanger, over which a fan blows outdoor air. Over time, water vapor from the outdoor air condenses on the (very cold) heat exchanger surface, then freezes, forming a layer of frost. This frost layer impedes heat transfer between the refrigerant and outdoor air, reducing the HP's heating capacity. To remedy this capacity reduction, HPs periodically switch to cooling mode, absorbing energy from the indoor air and driving hot refrigerant through the outdoor heat exchanger. While this reversal melts frost build-up on the outdoor heat exchanger, it also cools the indoor air. To avoid blowing cold air on occupants, air-source HPs use resistance heat during defrost cycles. 

Defrost cycles cause high current draws by simultaneously running the compressor, the fans, and especially the current-intensive resistance heat. During defrost cycles, the test house's HP historically selected Stage II (60 A) or III (80 A) of resistance heat regardless of the outdoor temperature. This behavior risked exceeding a 100 A whole-home current limit due to synchronization with other appliances. To mitigate this risk, the HP's settings were adjusted through its customer-facing user interface. In the updated settings, defrost cycles use only Stage I (40 A) when the outdoor temperature is above -9.4 $^\circ$C, Stage II between -9.4 $^\circ$C and -15 $^\circ$C, and Stage III below -15 $^\circ$C. 

Predicting defrost cycles is key to current-limiting control in homes with air-source HPs. Anticipating the onset of a defrost cycle and the magnitude of the associated current draw allows the control system to turn other devices down or off, or to delay the defrost cycle. A variety of methods exist to predict frost formation or defrost cycles \cite{MA2023127030}. However, these methods typically require detailed physical models of the outdoor heat exchanger or measurements of internal HP states such as the outdoor heat exchanger temperature or refrigerant flow rate. These models and measurements typically require manufacturer-level access to HP hardware. As this paper does not assume manufacturer-level hardware access, it instead develops a simple data-driven method for defrost prediction.

The data-driven method predicts defrost cycles indirectly by monitoring the supply air temperature $T_\text{sup}$ ($^\circ$C). As frost builds up, the supply air temperature decreases due to reduced heat transfer from the outdoor air to the refrigerant. The supply air temperature is predicted at time $k$ through linear autoregression:
\begin{equation}
\hat T_\text{sup}(k + 1) = \sum_{\ell=0}^{10} a_\ell T_\text{sup}(k - \ell) .
\end{equation}
The low-level controller predicts a defrost cycle if (a) the predicted supply air temperature $\hat T_\text{sup}(k+1)$ drops below the return air temperature at time $k$, and (b) the HP is on at time $k$. The stage of resistance heat used during defrost is predicted from the outdoor temperature. Once the prediction logic was finalized, the low-level controller correctly predicted the onset of all defrost cycles. It also occasionally predicted defrost cycles that did not happen. These false positives caused a few unnecessarily conservative control actions, but did not substantially degrade the overall control system performance.

\section{Results and discussion}
\label{testing_results}

The current-limiting control system described in Sections \ref{high_level_formulation} and \ref{low_level_controller} was tested from December 25th, 2023, to January 25th, 2024, in the test house described in Section \ref{testHouse}. Data from these 31 days were compared to baseline data collected in December of 2022 and January of 2023, when the house had no current-limiting control system. Both the testing and baseline periods included cold snaps, when outdoor temperatures fell below -20 $^\circ$C. The mean outdoor temperatures were -2.7 $^\circ$C over the testing period and 1.25 $^\circ$C over the baseline period. There were three occupants in the test house during the testing period and 1-2 occupants during the baseline period, as discussed in Section \ref{baselineOperation}. Due to lower outdoor temperatures and higher occupancy, the current-limiting control problem was more difficult during the testing period than it would have been during the baseline period.

\subsection{Control system behavior}

\begin{figure}[ht]
\centering
\includegraphics[width=0.45\textwidth]{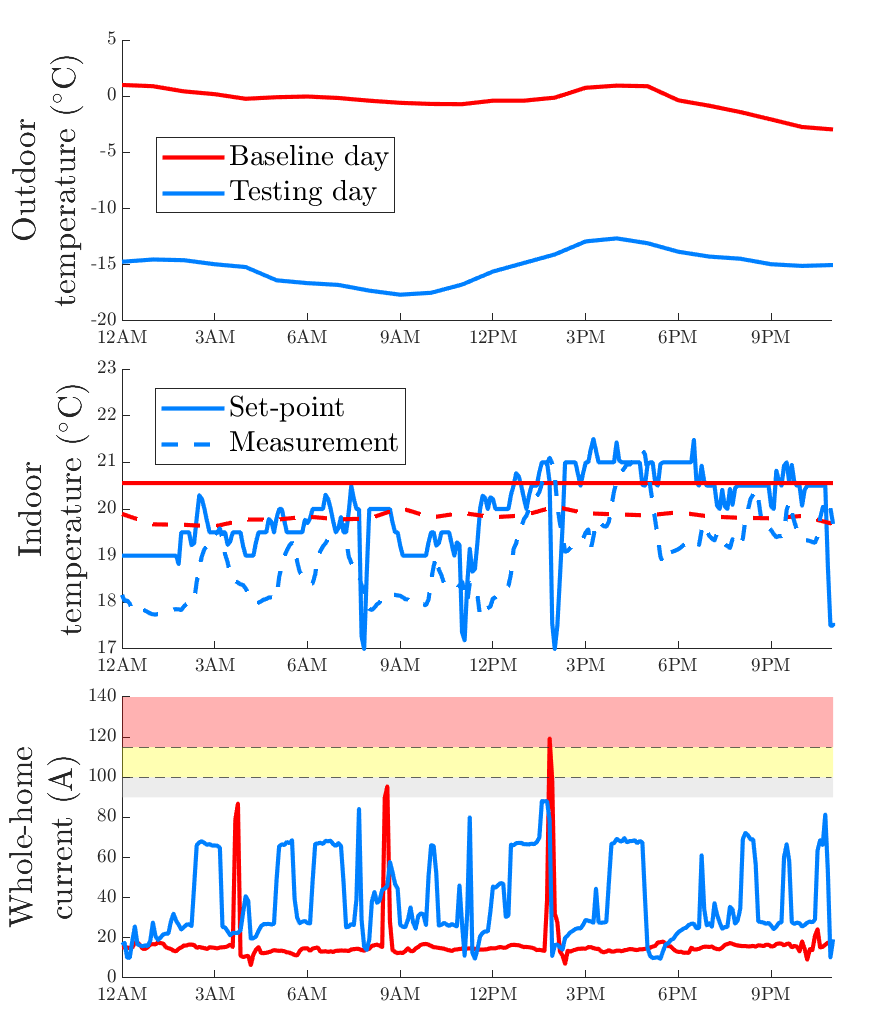}
\caption{The control system maintained peak whole-home current below 90 A on a challenging test day (blue curves), while current exceeded 120 A on a mild baseline day (red).}
\label{comparison_RBC_MPC1}
\end{figure}

Fig. \ref{comparison_RBC_MPC1} shows the control system behavior on two representative days: one in the baseline period (red curves) and one in the testing period (blue curves). The top plot shows that the outdoor temperature was much lower on the test day than on the baseline day, making current-limiting control more challenging. Despite this challenge, the current-limiting control system maintained the whole-home current (bottom plot) below 90 A. On the baseline day, by contrast, the current spiked to 120 A when a HP defrost cycle synchronized with WH operation due to a large water draw. 

The middle plots in Fig. \ref{comparison_RBC_MPC1} show the indoor temperature set-points (dashed curves) and measurements (solid curves). The red curves show that, as discussed in Section \ref{tracking}, the manufacturer's controls imperfectly tracked the set-point with a steady-state offset equal to the dead-band half-width, $\gamma = 0.5$ $^\circ$C. The blue curves show that the current-limiting control system adjusted the indoor temperature set-point quite dynamically. It lowered the set-point somewhat overnight to reduce the use of resistance heat, and preheated the home during the warmer afternoon, when the HP COP was highest. On three occasions, the low-level controller dropped the set-point sharply to  $\simm$17 $^\circ$C for about five minutes. This was done to smooth out recovery from defrost cycles, after which the HP often attempts to run the compressor at full speed while also using resistance heat.

\subsection{Current-limiting performance}

\begin{figure}[ht!]
\centering
\includegraphics[width=0.45\textwidth]{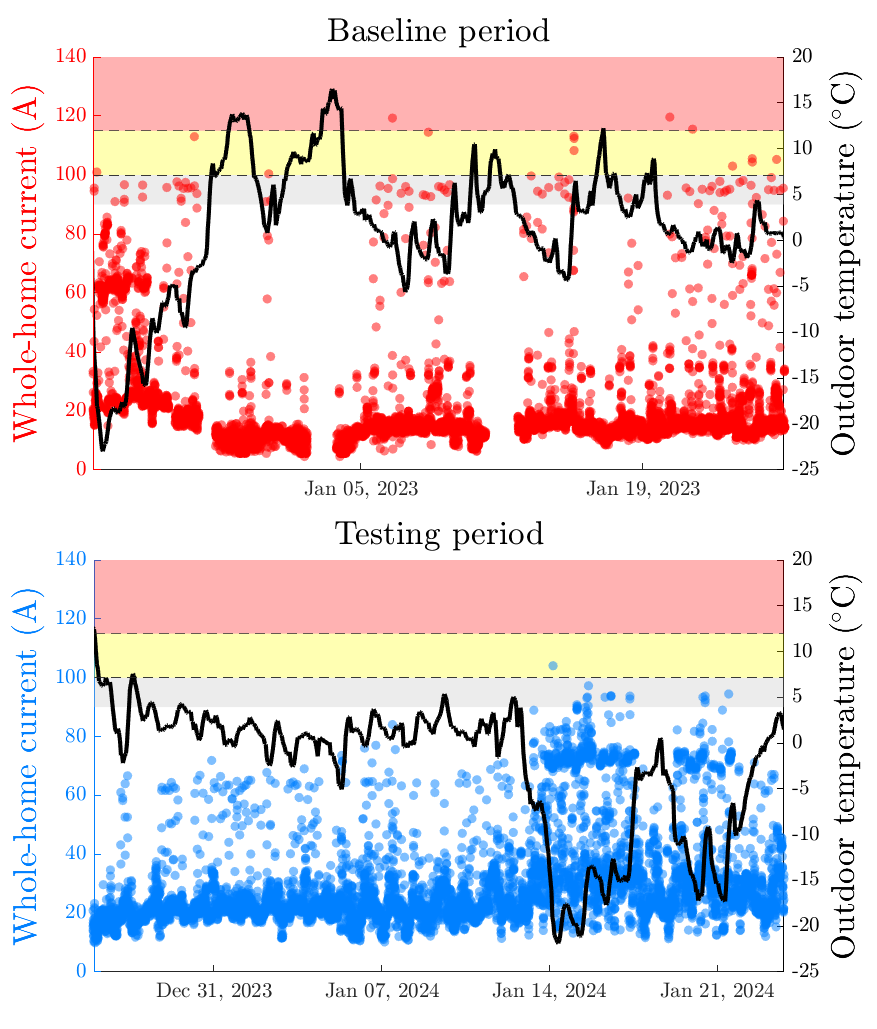}
\caption{Whole-home current (left axes) often exceeded the safe limits of a 100 A panel in the baseline period (top), but not in the testing period (bottom).}
\label{current_weather_timeseries}
\end{figure}

Fig. \ref{current_weather_timeseries} shows the whole-home current (left axes) and outdoor temperature (right axes) during the baseline period (top plot) and testing period (bottom plot). The black curves show the outdoor temperature, which fell below -20 $^\circ$C during cold snaps in both periods. During the baseline period, the whole-home current was often in the 80-100 A range and sometimes exceeded 115 A, where thermal trips are likely. In the testing period, by contrast, the whole-home current rarely exceeded 90 A (0.61 times per day, vs. 2.2 times per day in the baseline period), briefly exceeded 100 A only once, and never exceeded 105 A.

\begin{figure}[ht]
\centering
\includegraphics[width=0.45\textwidth]{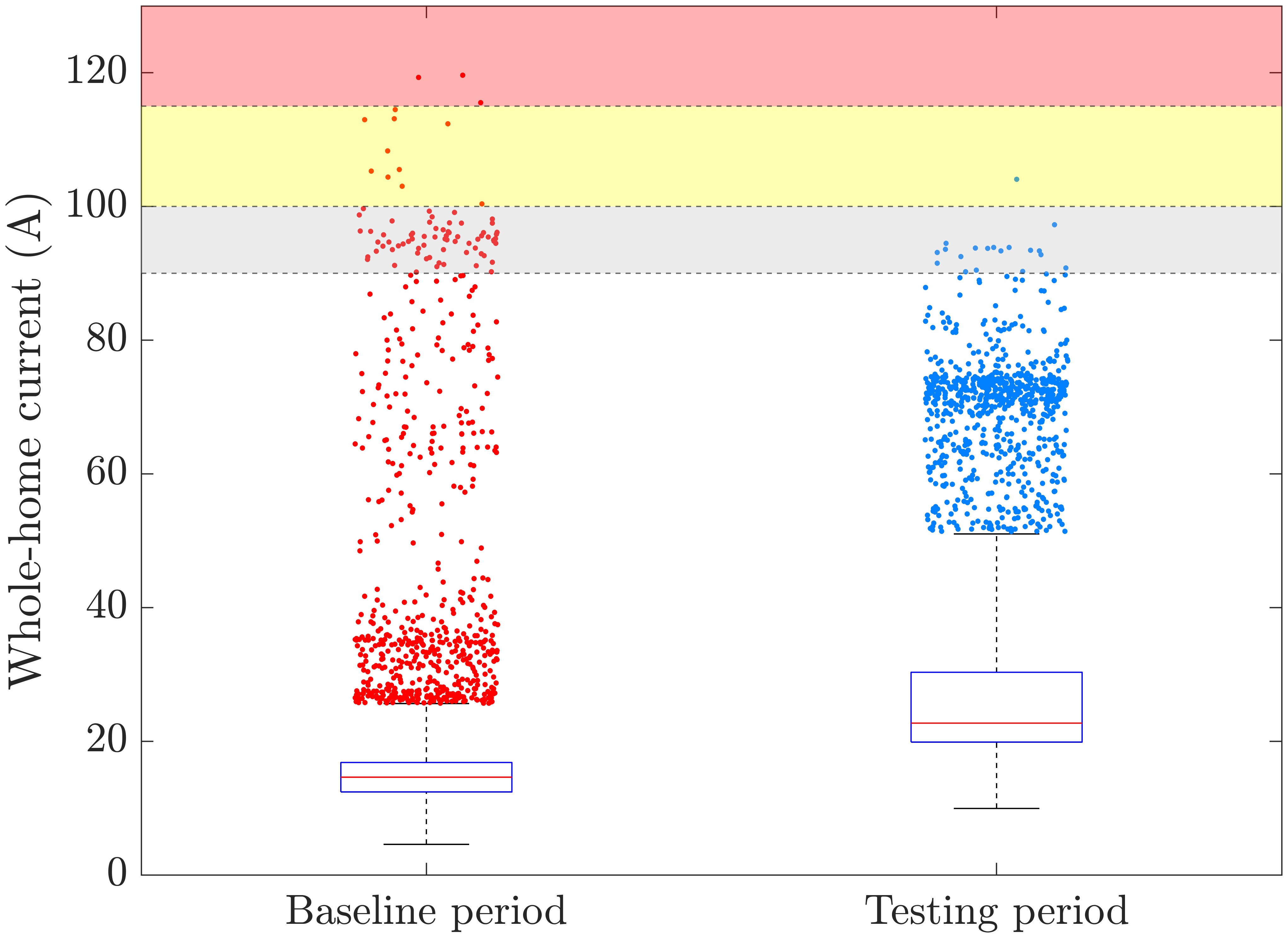}
\caption{The control system shifted the distribution of the whole-home current from an unsafe range in the baseline period (left) to a safe range in the testing period (right).}
\label{Trip Performance}
\end{figure}

Fig. \ref{Trip Performance} shows box-and-whisker plots of the whole-home current distributions during the baseline period (left) and testing period (right). In these plots, red lines are medians, blue boxes span the 25th to 75th percentiles, and dots (red for baseline, blue for advanced control) are outliers. Currents in the gray region risk thermal trips if another major appliance turns on. Currents in the yellow region risk thermal trips if sustained for ten minutes or longer. Even brief currents in the red region would likely cause thermal trips. While whole-home current often exceeded the safe limits of 100 A infrastructure during the baseline period, in the testing period it remained below 100 A for all but one five-minute time step. During that time step, the whole-home current reached 104 A due to a Wi-Fi communication delay that caused the WH to remain on during a HP defrost cycle. This example shows that robustness to communication delays and failures is an important area for control system improvement. While undesirable, a whole-home current of 104 A for five minutes is too mild and brief to pose significant risk of thermal tripping. The key message of Fig. \ref{Trip Performance}, therefore, is that the control system maintained the whole-home current at safe levels over the testing period for electrical panels and service rated at 100 A.

\subsection{Occupant comfort}
\label{occupantComfort}

The control system maintained occupant comfort while limiting whole-home current. The time-average indoor temperature was 20.2 $^\circ$C over the baseline period and 19.8 $^\circ$C over the testing period. The time-average supply water temperature was 47.8 $^\circ$C over the baseline period and 47.6 $^\circ$C over the testing period. These results suggest that the occupants' temperature preferences were met, a finding corroborated through observations during on-site visits and through occupant surveys.

\begin{figure}[ht]
\centering
\includegraphics[width=0.45\textwidth]{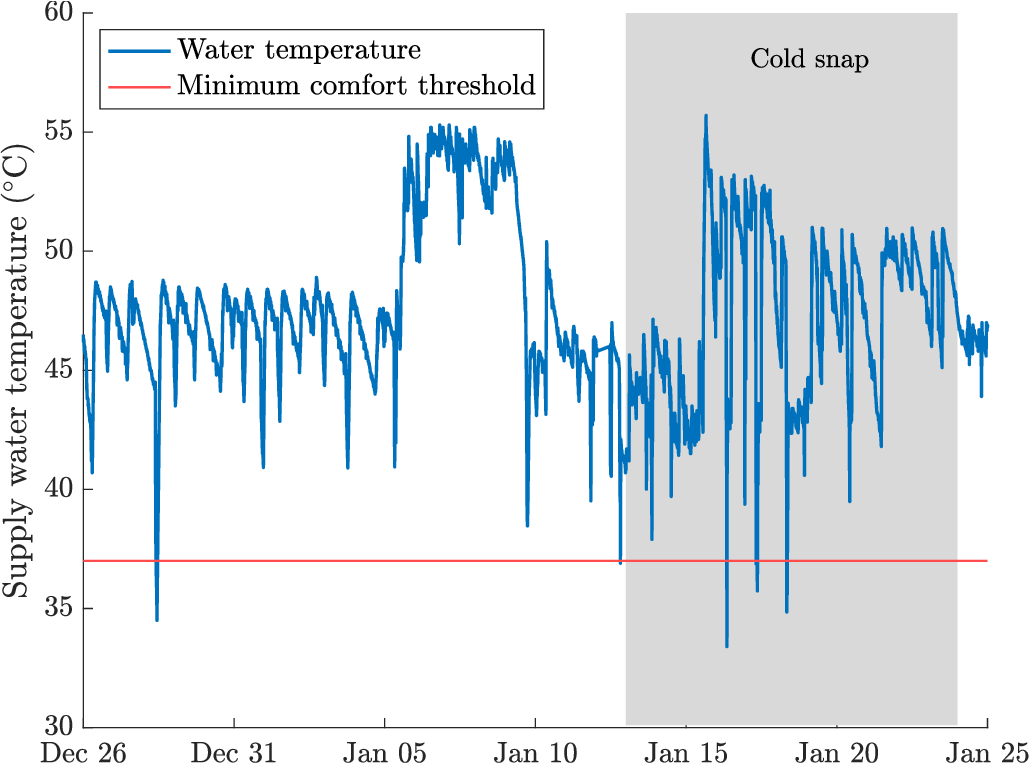}
\caption{Although supply water temperatures (blue curve) dropped below the comfort threshold  (red) four times over the testing period, occupants reported no discomfort.}
\label{WH Performance}
\end{figure}

Fig. \ref{WH Performance} shows the supply water temperature (blue curve) over the testing period. The red line at 36.6 $^\circ$C (roughly body temperature) is a minimum threshold below which occupants likely experience discomfort \cite{showerTemp}. Over the testing period, the supply water temperature briefly dropped below the minimum comfort threshold four times. On these four occasions, occupants drew lukewarm water for a total of four minutes. These temperature violations were due to back-to-back showers that coincided with the low-level controller turning off the WH after predicting a defrost cycle or reading a whole-home current above 90 A. To mitigate temperature violations, starting on January 15, the WH was preheated by sending a set-point of 54.4 $^\circ$C from noon to 2 PM and from 4 AM to 6 AM. Although this preheating strategy did not eliminate all temperature violations, the occupants did not report discomfort in periodic surveys. Without this preheating strategy, occupants would likely have drawn lukewarm water at times during the cold snap due to frequent defrost cycles and high use of backup heat.

\subsection{Accommodating a second electric vehicle}
\label{ev_sim_section}

The field experiments described above included a plug-in hybrid EV with Level I (1.8 kW) charging. As Level I charging draws relatively little current, it was left under occupant control and lumped in with the rest of the uncontrolled loads described in Section \ref{rest_of_house}. These experiments can be considered representative of the 41\% of U.S. households that have zero (8\%) or one (33\%) vehicle \cite{census2022}. However, 37\% of U.S. households have two vehicles and 22\% have three or more. Furthermore, U.S. households with EVs may install Level II charging stations, which typically provide 15-50 A at 240 V. To broaden the applicability of this paper's methods, this section presents simulations that include a second EV with a 70 kWh battery and a charging limit of 11.5 kW. 

House operation with a second EV is simulated by solving the MPC problem in Eqs. \eqref{eq:objective}-\eqref{eq:integer_variables} with additional terms representing the second EV, and with $S = 1$ scenario. Only high-level control is simulated; in practice, the low-level controller would add further robustness. The simulation uses historical weather and electrical load data from the house on a day when the time-average outdoor temperature was -15 $^\circ$C. The EV is modeled as in \cite{priyadarshan2024edgie}, begins with a 60\% full battery, drives typical distances at typical times for U.S. commutes, and charges only at home. The EV is restricted to on/off charging on 48 A and must have a full battery before the morning commute. To avoid frequently switching the EV charger on and off, the optimization allows at most 16 switches (or eight charging periods) over the simulation day:
\begin{equation}
\sum_{k=0}^{K-2} | z_\text{EV}(k+1) - z_\text{EV}(k) | \leq 16.
\end{equation}
The binary variable $z_\text{EV}(k)$ equals one if the EV charger is on at time $k$ and equals zero otherwise.

\begin{figure}[t]
\centering
\includegraphics[width=0.475\textwidth]{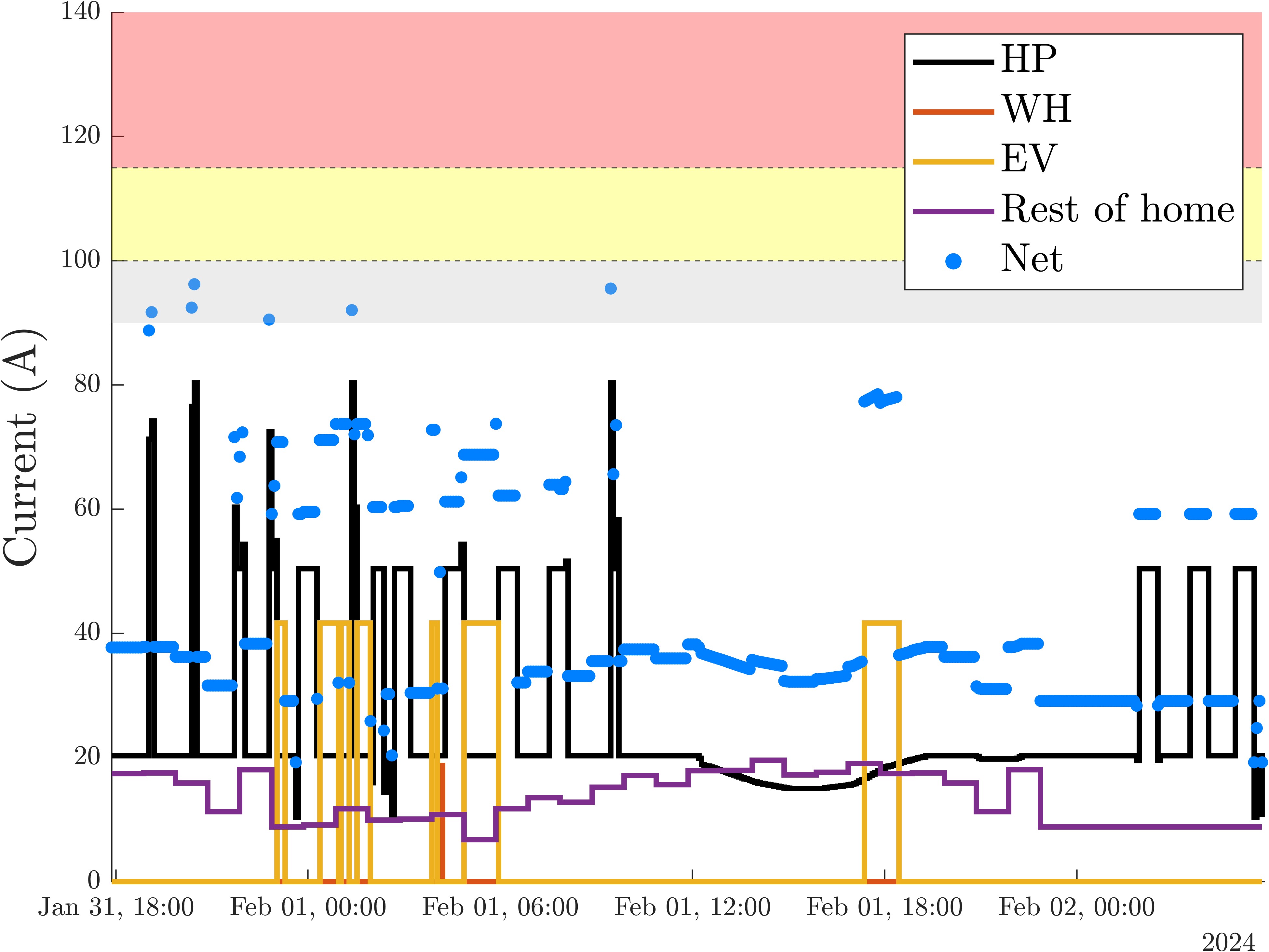}
\caption{Optimized whole home operation including an in-situ Level II EV charger}
\label{smart_whole_home_control}
\end{figure}

Fig. \ref{smart_whole_home_control} shows current draws over the simulated day. In Fig. \ref{smart_whole_home_control}, the purple `Rest of home' curve is the 99th percentile $q_\alpha$ of the uncontrolled current $I_u$. With a typical EV charging profile and no optimization, the whole-home current would exceed 150 A multiple times on this day due to coincident operation of the EV charger and other major appliances. This would certainly trip a 100 A breaker. With optimization, however, the whole-home current (blue dots) remains below 100 A at all times, even though the second EV draws 48 A when charging. The whole-home current is limited by interleaving the EV charging (yellow curve) with the HP operation (black curve). The HP turns off to make room for the EV within the 100 A limit. With this staggered charging profile, the EV battery gets filled before the morning commute. The space and water temperatures also remain in comfortable ranges. While these simulation results are not conclusive, they suggest that advanced control can maintain the current of an all-electric home with two EVs below the safe limits of panels and service rated at 100 A.

\section{Limitations and future work}
\label{future_work}

As the field experiments in this paper were the first of their kind, there are many opportunities to extend them. First, the control system could be tested in other climates or buildings. The experiments in this paper included one all-electric house with code-minimum insulation under outdoor temperatures as low as -20 $^\circ$C. Buildings in colder climate zones could require larger heat pumps or more resistance backup heat. Larger buildings, or buildings with worse insulation or air sealing, could also prove more challenging for current-limiting control. Testing under a wider set of conditions could build confidence that advanced control can enable electrification of older housing with few or no electrical upgrades. This testing could also generate data that enable economic analysis to identify settings in which current-limiting control provides an attractive return on investment.

Second, the control system could be tested with other devices. Notably, testing with a real second EV with Level II charging would buttress the simulation results in Section \ref{ev_sim_section}. While the simulation results suggest that an all-electric home could accommodate two EVs within a 100 A limit, proof in hardware would be more conclusive. The control system could also be tested with solar photovoltaics, stationary battery energy storage, thermal energy storage, or shiftable appliances such as dishwashers or clothes washers and dryers. Adding one or more of these devices should make current-limiting control easier, but would require updating control algorithms.

Third, the control system could be modified to improve robustness to communication delays or outages. This might involve connecting devices to a Home Area Network over a communication channel other than Wi-Fi, possibly the house's own electrical wiring \cite{aderibole2021power, aderibole2022distributed}. Alternatively or additionally, algorithms could be modified to fall back on reasonable heuristics after communication failures.

Fourth, the control system could be modified to reduce computational intensity. As formulated in Eqs. \eqref{eq:objective}-\eqref{eq:integer_variables}, the optimization problem that the high-level controller solves every five minutes is a mixed-integer convex program with on the order of 10$^4$ variables. This problem would be challenging to solve in real time on a low-cost micro-controller. A number of strategies might reduce computational complexity, such as warm-starting the solver with the last time step's solution, removing dependence on the modeling language CVX through automated code generation as in \cite{mattingley2012cvxgen}, reducing the problem dimension by using longer time steps further ahead in the prediction horizon, or eliminating the dependence on multiple forecast scenarios through improved modeling of conditional water draw distributions. Alternatively, predictive load-shifting heuristics could be added to the real-time control layer, enabling emulation of the high-level MPC behavior without the need to solve optimization problems online. Such heuristics, if effective, could significantly lower barriers to implementation at scale.

Fifth, the control system could be modified to engage more directly with occupants. The control design in this paper limited occupant input to specifying temperature preferences for space conditioning and water heating. However, related work such as \cite{kim2022mysmarte} has shown that engaging occupants more actively, for example by providing tips or coordinating games between neighbors, can improve the user experience and deepen performance improvements from advanced control.

Sixth, this paper developed a control system that protects electrical infrastructure at the whole-building level, but electrification can also stress infrastructure at smaller and larger scales. Future work could develop control systems that maintain currents below safe limits at the smaller scale of individual circuits within a building. At larger scales, load-shifting control systems could also protect the distribution transformers that typically serve two to 10 homes, or the substation transformers that serve whole neighborhoods. Smaller- and larger-scale current-limiting control systems could further lower barriers to electrification, but might require different control architectures or methodologies.

Seventh, other supervisory control methodologies could be investigated. The MPC implementation in this paper required the offline training of a thermal circuit model of the building temperature dynamics, as well as a time-series forecasting model for disturbance prediction. These steps were performed once and the resulting models were used without changes throughout the testing period. Alternatively, model structures or parameters could be updated adaptively to account for changes in occupant behavior, for equipment degradation, and for other time-varying effects. Model-free control techniques, such as reinforcement learning \cite{WANG2025124328_DRL} and data-enabled predictive control \cite{chance_DeePC}, could also be investigated. These approaches could reduce the deployment effort required for modeling and machine learning, potentially improving scalability.

Eighth, and perhaps most importantly, compliance with electrical codes for sizing electrical panels and service could be investigated. As far as the authors are aware, the National Electrical Code \cite{nationalCode} currently recognizes the capability of control systems such as the one developed in this paper to protect electrical infrastructure by coordinating device operation. As a precursor to widespread deployment of systems such as these, however, implementation details would need to be clarified and perhaps updated to more explicitly accommodate software-based approaches to current limiting.

\section{Conclusion}

This paper presented the first field demonstration that advanced control can enable older houses to fully electrify without upgrading electrical panels or service. This paper developed a two-level control system that adjusts the temperature set-points of a heat pump and water heater. The control system was tested in an all-electric house in Indiana over 31 days, including a cold snap when outdoor temperatures dropped below -20 $^\circ$C. The control system maintained the whole-home current below the safe limits of electrical panels and service rated at 100 A. If electrical codes allow, current-limiting control could save owners of older homes on the order of \$2,000 to \$10,000 in electrical work when electrifying. Avoiding these costs could lower a significant barrier to residential electrification, accelerating emission reductions and improving access for disadvantaged communities.

\section*{CRediT authorship contribution statement}

{\bf Elias Pergantis:} Conceptualization, Methodology, Investigation, Formal Analysis, Data Curation, Visualization, Writing – Original Draft, Writing – Review \& Editing. {\bf Levi D. Reyes Premer:} Software, Methodology, Formal Analysis, Data Curation, Visualization, Writing – Original Draft, Writing – Review \& Editing. {\bf Alex H. Lee:} Software, Visualization, Writing – Original Draft, Writing – Review \& Editing. {\bf Priyadarshan:} Software, Writing – Original Draft. {\bf Haotian Liu:} Review \& Editing, Project Administration, Funding Acquisition. {\bf Eckhard Groll:} Review \& Editing, Project Administration. {\bf Davide Zivani:} Review \& Editing, Project Administration, Funding Acquisition. {\bf Kevin Kircher:} Conceptualization, Methodology, Writing - Review \& Editing, Project Administration, Funding Acquisition.

\section*{Declaration of competing interest}

The authors declare that they have no known competing financial interests or personal relationships that could have appeared to influence the work reported in this paper.

\section*{Data availability}

Data will be made available on request.

\section*{Acknowledgments}

The Center for High-Performance Buildings at Purdue University supported this work (project CHPB-26-2023). E. Pergantis was also supported by the Onassis Foundation as one of its scholars, as well as the American Society of Heating and Refrigeration Engineers (ASHRAE) through the Grant-in-Aid award. The authors would like to thank the occupants of the test house for their patience and cooperation during testing.

%\appendix
\section*{Appendix. Acronyms and notation}
\label{notation}

This paper used eight acronyms: heat pump (HP), water heater (WH), electric vehicle (EV), United States (U.S.), model predictive control (MPC), Electric Power Research Institute (EPRI), coefficient of performance (COP), and support vector machine (SVM). Table \ref{notationTable} summarizes the mathematical symbols used in this paper. These symbols combine with the following subscripts and superscripts: $m$ (thermal mass), `out' (outdoor), `OL' (open loop), $r$ (resistance), $w$ (water), $s$ (scenario index), $e$ (electricity), $t$ (thermal comfort), $a$ (ambient air), $u$ (uncontrolled), and `sup' (supply). Variables without subscripts usually refer to the HP. Variables with hats, such as $\hat w$, denote predictions. Variables with tildes, such as $\tilde T$, denote set-points. Variables with daggers, such as $T^\dagger$, denote user preferences. Underlined and overlined variables, such as $\underline P$ and $\overline P$, denote lower and upper bounds.

\begin{table}[ht!]
\caption{Mathematical notation} 
\label{notationTable} 
\centering
\small

\begin{tabular}{ l | l }
Symbol (Units) & Meaning \\
\hline
$t$ (h) & Continuous time \\
$\Delta t$ (h) & Time step duration \\
$k$, $\ell$ (-) & Discrete time indices \\
$K$ (-) & Number of time steps \\
$\mathcal K$ (-) & Set of times \\
$s$ (-) & Scenario index \\
$S$ (-) & Number of scenarios \\
$\mathcal S$ (-) & Set of scenarios \\
$T$ ($^\circ$C) & Temperature \\
$\theta$ ($^\circ$C) & Boundary temperature \\
$\eta$ (-) & Coefficient of performance \\
$P$ (kW) & Electric power \\
$I$ (A) & Current \\
$V$ (V) & Voltage \\
$f$ (-) & Power factor \\
$R$ ($^\circ$C/kW) & Thermal resistance \\
$C$ (kWh/$^\circ$C) & Thermal capacitance \\
$\tau$ (h) & Tracking time constant \\
$\gamma$ ($^\circ$C) & Dead-band half-width\\
$a$, $b$ (-) & Dynamics parameters \\
$\dot Q$ (kW) & HP thermal power \\
$w$ (kW) & Exogenous thermal power \\
$e$ (kW) & Error in predicting $w$ \\
$\lambda$ (-) & $\hat w$ pessimism parameter \\
$\sigma$ (kW) & $\hat w$ standard deviation \\
$\beta$ (-) & Autoregressive parameter \\
$M$ (-) & Autoregressive memory \\
$X$ (mixed) & Autoregressive data matrix \\
$y$ (kW) & Autoregression target \\
$\pi$ (mixed) & Price \\
$z$ (-) & Binary or integer variable \\
\end{tabular}
\end{table}

% bibliography
\bibliographystyle{elsarticle-num}
\bibliography{plim.bib}

\end{document}